\documentclass{aa}
\usepackage[varg]{txfonts}
\usepackage{graphicx}
\usepackage{xcolor}
\usepackage{natbib}
\usepackage{rotating}
\usepackage{hyperref}

\defcitealias{Kuhnetal14}{K14}
\defcitealias{Gonzalezetal21}{Paper I}
\defcitealias{Getmanetal18}{G18}
\defcitealias{Joncouretal18}{J18}
\defcitealias{Gutermuthetal09}{G09}

\begin{document}
\title{S2D2: Small-scale Significant substructure DBSCAN Detection}
\subtitle{ II. Tracing episodes and gradients of star formation activity.}

\author{Marta Gonz\'{a}lez \inst{\ref{inst1}} \and Isabelle Joncour \inst{\ref{inst2}}  \and Estelle Moraux \inst{\ref{inst2}}  \and Fr\'{e}d\'{e}rique Motte  \inst{\ref{inst2}} \and Elisa Nespoli \inst{\ref{inst1}} \and Fabien Louvet \inst{\ref{inst2}} \and Maxime Valeille-Manet \inst{\ref{inst2},\ref{inst3}} \and  Vicent Mart\'{i}nez-Badenes \inst{\ref{inst1}} } 

\institute{{Universidad Internacional de Valencia (VIU),
C/Pintor Sorolla 21, E-46002 Valencia, Spain}\label{inst1}\and Univ. Grenoble Alpes, CNRS, IPAG, 38000 Grenoble, France\label{inst2} \and Laboratoire d’astrophysique de Bordeaux, Univ. Bordeaux, CNRS, B18N, all.e Geoffroy Saint-Hilaire, 33615 Pessac, France \label{inst3}}

\bibpunct{(}{)}{;}{a}{}{,} 

\abstract {The spatial analysis of young stellar objects has proved very valuable to describe and analyse star-forming regions and understand the star formation process. } 
{ Provide the community with a homogeneous catalogue of small, significant substructures (henceforth NESTs) extracted from the spatial distribution of Young Stellar Objects (YSOs) in a large, consistent sample of star-forming regions.  {Explore the relevance of small scale spatial substructure. and discuss the interpretation of NESTs as tracers of star formation activity and remnants of the star formation process. } }
{We apply our procedure to consistent catalogues of YSOs to obtain NESTs in a sample of star-forming regions. We apply a photometric classification scheme to obtain the evolutionary stage of YSOs and statistically explore the distribution of class 0/I objects as a proxy of recent star formation activity.}
{The region sample is diverse (in distance, size, structure, and global evolutionary stage), and we consequently find different structural properties and star formation histories. Most
NESTs in regions with high recent star formation activity show even higher levels of activity. 
Moreover, the proportion of NESTs with higher activity than the region average increases with the global level of activity of the region. In approximately half of the regions we also find significant spans in the evolutionary stages of the NESTs, consistent with gradients and episodes of star formation.}
{The combination of NESTs with a statistical exploration of the star formation history within each region provides robust and powerful insights into the star formation process. 
Our results support the role of NESTs as pristine remnants of star formation in highly active regions,   {stressing the role of fragmentation. The combination of small structures with large scale spatio-evolutionary patterns suggests hyerarchical, prolonged, dynamic, and complex star formation scenarios }}

\maketitle
\section{Introduction} 

Star formation (SF) is a complex, widespread phenomenon evolving at all spatial and time scales due to the rapid variation of diverse environmental factors. These include cloud density and dynamics, and, as SF proceeds, ionization, heating, and stellar winds, which become particularly significant when massive stars form. Thus, star-forming regions are characterised by intricate spatial patterns across a wide range of scales both in the cloud component and the newly formed Young Stellar Objects (YSOs). 
This work focuses on the spatial structure of the young object distribution, which shows a hierarchy of structures at scales from individual objects to clusters, complexes, and super-complexes. 

Spatial analysis of YSOs is a powerful tool to study the formation and evolution of star-forming regions, particularly when combined with age or evolutionary estimates. 
YSOs can be classified in different evolutionary stages, according to the infrared (IR) signatures linked to the accretion process. 
\begin{itemize}
\item{\textbf{Class 0/ I}: Less evolved objects, characterised by significant IR emission consistent with an accretion envelope with some emission from the protostar and a dense accretion disk in the Class I case.}
\item{\textbf{Class II}: YSOs in an intermediate evolutionary state, where the envelope has disappeared, showing photometric IR excess consistent with an accretion disk. }
\item{\textbf{Class III}: Evolved YSOs, showing very low or no IR excess from an evolved, dissipated disk.}
\end{itemize}
  
 The spatial distribution of YSOs according to their classes or evolutionary stages has long been used to study the evolution of SF within a region \citep[see e.g.][for the seminal study in NGC2264]{Sungetal09}.
The separation of areas of interest is key in such studies, and is often related to the retrieval of relevant dense substructures. Methods of substructure retrieval often leave out the smaller spatial scales due to resolution limitations and the difficulty of separating them from spurious or random effects.  {This work focuses specifically on small scales, complementing more traditional structure retrieval studies with their small-scales counterparts.}  

\citealt{Joncouretal18} (henceforth \citetalias{Joncouretal18}) analysed the small-scale substructures in the YSO distribution from Taurus, ensuring the significance by comparing the sample to complete spatial randomness (CSR). The retrieved structures, NESTs (from Nested Elementary STructures), were on an intermediate spatial scale between ultra-wide pairs and loose groups and their analysis proved very interesting. NESTs were typically associated with the molecular cloud, elongated, and aligned with the main gas filaments. Notably, approximately half of the NESTs (comprising less than 30\% of all YSOs) contained 3/4 of all the Class 0/I objects. The work of 
\citetalias{Joncouretal18} suggested that NESTs were the pristine remnants of the smallest collapsing gas structures. 

 {
In \citep{Gonzalezetal21} (henceforth \citetalias{Gonzalezetal21}) we automated some steps in \citetalias{Joncouretal18} for direct and consistent application in different regions, defining  S2D2 (Small-scale, Significant DBSCAN Detection).  S2D2 is a robust procedure for the detection of small substructures, enforcing that the retrieved clusters \citepalias[that we call NESTs as in][]{Joncouretal18} are significant compared to CSR.  S2D2 reaches the smaller scales and avoids random fluctuations by requiring strict statistical density criteria for parameter selection, which implies that it is not sensitive to large scale, moderate overdensities.  We detail the procedure, highlighting its strengths and limitations, in section \ref{s2d2}.}

We follow \citetalias{Joncouretal18} and \citetalias{Gonzalezetal21} and apply S2D2 to a large sample of star-forming regions where all the members were determined homogeneously. We combine the NESTs with evolutionary stage estimates of YSOs to determine the range where NESTs can be considered the spatial imprints of recent SF and their potential as tracers of the SF history within a region.

This work has 6 sections, the first one being this introduction. In section \ref{method} we describe the samples and methods applied. We have split the results section into two: Section \ref{results_general} shows aggregated results highlighting general trends, while in section \ref{results_spatial} we study the spatial distribution of NESTs. In section \ref{discussion} we contextualise and explore the role of NESTs as pristine remnants of SF and tracers of the SF history, and finally, in section \ref{conclusions} we summarise our results and conclusions. Additionally, in Appenddix \ref{appendixCaracts} we provide some useful tables (with the characteristics of all regions, acronyms used in this work and a list of the online complementary material), and in Appendix  \ref{appendixRegions} we detail the results for each of the regions analysed.

\section{Method}\label{method}
In this section, we describe the sample and procedures applied to construct the catalogue of significant substructures that will be the base to compare SF amongst regions.
Both the use of samples homogeneously obtained and a robust, common method of retrieval are important to ensure that the comparison of substructures retrieved amongst regions is as fair and unbiased as possible.

\subsection{Star-forming region sample: the MYStIX and SFiNCs programs}

The MYStIX program \citep[][]{Feigelsonetal13} was an observational project designed to obtain the young stellar population of a sample of massive star-forming regions within 4 kpc distance from the Sun. 
The program complemented near-infrared data with X-ray observations (allowing for the detection of a larger number of objects) and curated a list of member candidates for each region trying to minimise contamination. 
The final sample of probable members of each region was obtained using Bayesian inference \citep{Broosetal13}
combining data from \textit{Spitzer} IRAC \citep{Fazioetal04}, 2MASS \citep{Skrutskieetal06}, UKIDSS \citep{Lawrenceetal07}), and \textit{Chandra} \citep{Weisskopfetal00}. Known OB stars from the literature were also added to the catalogues. 

The SFiNCs program \citep{Getmanetal17} was a subsequent program aimed at extending and complementing the MYStIX sample and focused specifically on smaller, nearby regions. The observational advantages of closer regions allowed for a simpler membership and evolutionary stage determination, but the project was designed and implemented to keep sample consistency between programs. 

These projects provided the community with homogeneous and comparable catalogues young stellar of members of all the studied regions \citep[][]{Broosetal13, Povichetal13, Getmanetal17}, suited to spatial structure studies.  {Despite their age, the MYStIX and SFiNCs catalogs are still relevant and some of the most complete samples of YSO candidates in their respective regions, due to the inclusion of selected X-ray sources.  We verified this comparing the samples with  the SPICY catalog,  a large YSO candidate catalog based on IR-excess containing more than 100.000 sources \citep[][]{Kuhnetal21Spicy}.  We only found matches in approximately one third of the regions.  In  all them,  the number of  SPICY members is less than half the number of MYStIX and SFiNCs candidates,  and most of them are already in the MYStIX/SFiNCs sample.}

\citetalias{Kuhnetal14} applied a parametric mixture model of isothermal ellipsoids to detect the substructures present in each of the MYStIX regions, and the same methodology was later extended to the SFiNCs sample \citep[][henceforth \citetalias{Getmanetal18}]{Getmanetal18}. These subclusters were used to describe the regions, and their physical, kinematical, and age characteristics were estimated in later studies \citep{Kuhnetal15, Kuhnetal19Kinematics, Getmanetal18AgeGradients}. 

In this work, we apply S2D2 to retrieve the NESTs in the MYStIX and SFiNCs catalogues, providing a complementary perspective to that from \citetalias{Kuhnetal14} and \citetalias{Getmanetal18} analysis.

\subsection{S2D2: NEST retrieval and calibration}\label{s2d2}
We now briefly describe the procedure S2D2, used to retrieve the NESTs that constitute our catalogue, described in Appendix \ref{appendixCaracts}. We refer the readers to \citetalias{Gonzalezetal21} for an extensive description of the method, its calibration and behaviour in both synthetic and observed control regions, as well as to the available public implementations \footnote{\url{https://starformmapper.org/algorithms/}}.  

S2D2 stands for Small-scale Significant DBSCAN Detection, and  {selects the parameters of  DBSCAN \citep{Esteretal96} using statistical criteria to detect small-scale structures that are significant above random expectation. DBSCAN is a well known density based algorithm based on two parameters:  The scale $\varepsilon$, which defines a local neighbourhood for each star,  and a minimum number of neighbours $N_{min}$.  Together,  $\varepsilon$ and $N_{min}$ define a nominal density requirement $\rho_{nom}=\frac{N_{min}}{\pi \varepsilon^2}$.}
 {DBSCAN does not force all the elements of the sample into clusters, or a specific number of structures and can detect structures of any shape. However, its parameter choice limits its output to a specific density, and it can be weak to gradual density variations, which may lead to spurious separation or merging of structures.}

 {Appendix C in \citetalias{Joncouretal18} compares DBSCAN with other clustering methods, but many variations have appeared to improve it in terms of efficiency, parameter dependence, and quality of clusters \citep{BusraYi21}.}
  {Given that S2D2 specifically addresses parameter selection, and that the scope of the method are datasets of moderate size, the most relevant category is the quality of clusters, particularly in datasets characterised by large density ranges such as star-forming regions. Despite being beyond the single scale focus of this work, the multiscale variations OPTICS and HDBSCAN, that probe a range in $\varepsilon$ and produce a hierarchy of substructures, are relevant in a star formation context. \citet{Canovasetal19} found comparable results for the three algorithms in $\rho$ Ophiuchi, but often HDBSCAN is favored over OPTICS due to computational speed and because cluster selection is direct. }
 { In addition to being able to consider datasets with density variations, the high sensitivity of  HDBSCAN justified its selection by \citealt{HuntReffert21} to retrieve clusters from \textit{Gaia} data. The disadvantage of HDBSCAN is a false positive problem which forces a curation a posteriori of the structures. }
 {In some of the tests from \citealt{BusraYi21} HDBSCAN retrieved a larger number of clusters than expected, pointing to the splitting of single structures which did not appear for DBSCAN. The work from \citealt{HuntReffert21} also showed a very good precision for DBSCAN, with low levels of false positives. The counterpart was a decreased sensitivity, with DBSCAN failing to detect some structures. The false positive issue is particularly problematic at small scales, which are severely affected by random fluctuations. This justifies the choice of DBSCAN for our work, even at the cost of completeness. }

 { The procedure S2D2 selects the parameters of DBSCAN to ensure a theoretical level of significance compared to a random distribution. }
The retrieval scale $\varepsilon$ is chosen using the smallest scale that shows a transition from an excess to a defect of stars in the region with respect to (w.r.t.) random expectation, as shown by the one-point correlation function  $\Psi$  \citep[OPCF, described in][]{Joncouretal17}.
We calculate $N_{\rm{min}}$ using the probability density function of the $n^{th}$ nearest neighbour \citepalias[][]{Joncouretal18} to guarantee that the structures retrieved have a significance above 99.75 \% w.r.t. random.  The density of the CSR control distribution $\rho_{\rm{CSR}}$ representative of the complete region is calculated using the expression described in \citetalias{Gonzalezetal21}.

  {
In \citetalias{Gonzalezetal21}, we calibrated S2D2 with 90 synthetic regions (10 random realizations of 9 different spatial distributions). The regions followed distributions representing clusters with global fractal substructure (fractal-box models) and large-scale concentrations (radial or Plummer distributions) with various parameters, including fractal and radial approximations of CSR. We found low levels of spurious detection in CSR,  and the number of NESTs (as well as the fraction of objects within them) increased with the level of substructure or concentration. 
In substructured, fractal distributions we retrieved small NESTs all over the region, with a limited fraction of objects within the NESTs.  On concentrated distributions we often found groups of NESTs corresponding to a single large-scale structure and comprising a substantial proportion of YSOs.}

{We also applied S2D2 to observed regions,  beyond the box-fractal / radial paradigm, to evaluate the performance of the algorithm in real star forming regions. We analysed an updated sample of Taurus, with $\sim 30 \%$ more members than  \citetalias{Joncouretal18}, and obtained a similar number, position and size of NESTs.  This supported the reliability of the automations in the process and the robustness of the structures retrieved by S2D2.  }
 {In  the Carina region,  using the sample of MYStIX also in this work, we recovered NESTs in the areas corresponding to known clusters Trumpler 15, 14, 16 and the Treasure Chest but not Bochum 11, in the South.  NESTs were inside or close to the central ellipsoids of the structures retrieved by \citetalias{Kuhnetal14}, and all the members of NESTs in Carina were members of \citetalias{Kuhnetal14} clusters.  We also found that NESTs were always in zones of high clustering tendencies according to INDICATE \citep[a density-based statistical index that quantifies the degree of spatial clustering or association of each object][]{Buckneretal19}.   A substantial fraction of structures from \citetalias{Kuhnetal14} in Carina did not have NEST counterparts.  Some of them were in zones of significant INDICATE clustering tendencies and known clusters,  and are possibly large structures with densities that S2D2 is insensitive to. Others, however, have low density values and clustering tendencies and may be background density fluctuations.}

To further ensure that NEST retrieval is not affected by binaries and chance alignments, we have merged all the stars that are closer than a specified threshold. We have chosen an angular threshold of $2 ^{"}$ , similar to the resolution of Spitzer- IRAC and 2MASS observations, instead of a physical one because the distance span between regions is significant and a limit valid for the farthest regions can excessively degrade the closest ones. This implies that in the farthest regions, we will not be able to sample spatial scales as small as in the closer ones. 

 {\citetalias{Gonzalezetal21} analysed the results of S2D2, finding low levels of spurious detection and coherent behavior in a variety of situations, and supporting our confidence in NESTs. The appearance of groups of NESTs, while out of the initial scope, was consistent along both synthetic and real regions, and provides a way to trace significant large scale concentrations. However, we need to keep in mind that NESTs cannot produce a complete description of all the substructure within a region, as S2D2 is insensitive to structures of lower relative density.}

\subsection{Evolutionary state of YSOs, NESTs, and regions}\label{methodEvol}

This section describes the procedure we used to derive evolutionary stage estimates for each NEST. We use NGC2264 \citep[a region long studied with spatial analysis of objects of different evolutionary stages by e.g.][]{Sungetal09, Rapson14, Venuttietal18, Nonyetal21} as a benchmark to evaluate the validity and performance of our approach. In our tests, we obtain that our procedure provides a consistent global picture for NGC2264 even when we use different samples or evolutionary stage estimates. Detailed results of these tests can be accessed as online complementary material, as indicated in Appendix \ref{appendixCaracts}.

\subsubsection{Evolutionary stage estimates of YSOs} \label{estimEvol}
All the objects in the catalogues from \citetalias{Kuhnetal14} and \citetalias{Getmanetal18} were selected as young stellar members of the regions, based on photometric, statistical, and spatial criteria. For a fraction of these members, estimates of their evolutionary stage were provided.  
While both relied on IR photometry, the method to estimate the evolutionary stage of YSOs in the MYStIX regions differs from the one applied for SFiNCs.
The evolutionary stage classification of YSOs in MYStIX regions was performed by \citet{Povichetal13}, and relied on Spectral Energy Distribution (SED) fitting to calculate the $\alpha$ IR spectral slope index. For the closer, less crowded, SFiNCs regions \citet{Getmanetal17} provided a simpler classification of members with/without disks based on IR excess. 

For the sake of consistency and homogeneity, we have reclassified all the objects using the method by \cite{Gutermuthetal09} (henceforth \citetalias{Gutermuthetal09}), also applied by \citet{Rapson14} to NGC2264. \citet{Buckneretal19} found that within the objects in common, \citet{Rapson14} had more unambiguously classified sources than \citetalias{Kuhnetal14}. 

We briefly describe the classification scheme from \citetalias{Gutermuthetal09}, which uses Spitzer and 2MASS data to retrieve YSOs based on their IR excess. 
The method works in three steps \citepalias[phases I, II, and III, following the terminology in ][]{Gutermuthetal09}, and each of them applies some cuts in colour-colour and colour-magnitude diagrams to discriminate  YSOs from non-stellar contaminants. 
Phase I uses the Spitzer-IRAC 4-band data to characterise Class 0/I and Class II objects separating them from broad-band AGNs, and shock/PAH-emission sources.
In Phase II, objects without IRAC data of enough quality are characterised using dereddened 2MASS HJK$_s$ photometry. Phase III performs a final pass through the catalogue to retrieve some additional Class 0/I objects with Spitzer-MIPS 24-micron data. It also allows for the distinction of Class III from Transition Disk objects and more evolved stars and confirm the YSOs detected in previous phases. 
 
 The MYStIX and SFINCs datasets do not have MIPS data, and we obtain a very low number of coincidences when cross-matching the YSO catalogue with MIPSGAL sources. Thus, we did not include this information as it would not statistically improve the procedure, and could introduce differences in the classifications between regions. We note, however, that without the information from MIPS, we cannot perform phase III and objects classified as Class III cannot be distinguished from main-sequence stars. All the sources in this work were already selected as young stellar members of the regions by \citetalias{Kuhnetal14} and \citetalias{Getmanetal18}, so we assume that they are Class III.

Some objects in the catalogues do not have photometric data of sufficient quality to provide a reliable classification, and others are classified as non-stellar (e.g. AGN, PAH...) in the different phases. All these are discarded from our evolutionary analysis, while the rest will be kept to statistically evaluate the evolutionary status of the region and its NESTs. Each selected object is labelled Class 0/I, Class II, and  Class III. This classification, along with the NEST membership on each object, is provided as complementary material. 

 \subsubsection{KDE and Relative risk}\label{SectRelRisk}
 
 \begin{figure*}[ht!]
\begin{center}
\includegraphics[width=0.8\textwidth]{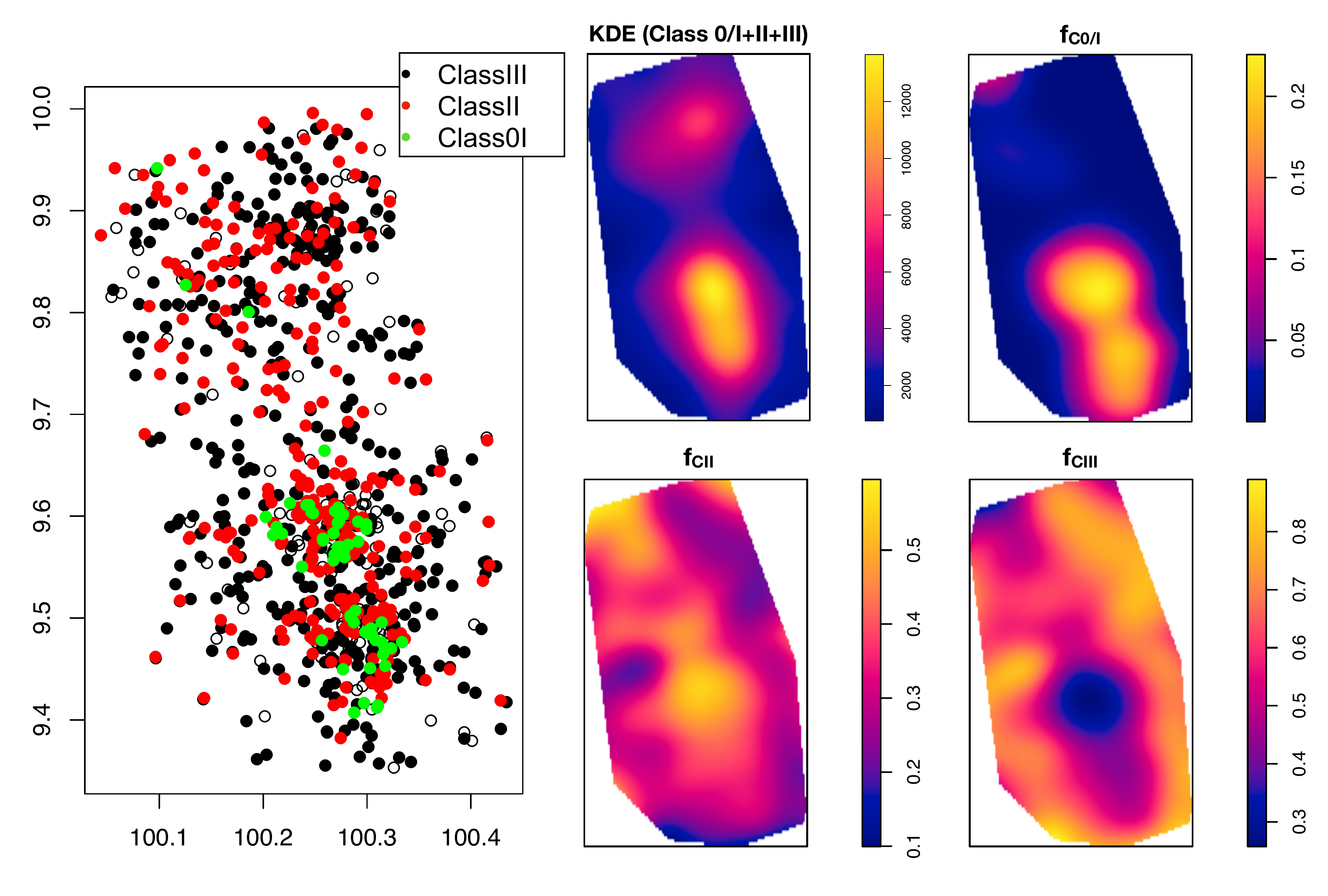}
			
\end{center}
\caption{Spatial distribution and relative risk maps for NGC2264. Left: Spatial distribution of objects in NGC2264. Dots represent each source in the complete sample, coloured according to the evolutionary stage classification described in section \ref{estimEvol} (green for Class0I, red for Class II, black for Class III, and empty for unclassified and non stellar). Right: Four panel composite showcasing the associated densities and relative risk maps.Top left: KDE estimate corresponding to the classified subsample in the left pannel. Top right: Relative risk maps of Class 0/I objects. Bottom panels: Relative risk maps of Class II and Class III objects.}
\label{relRisk}
\end{figure*}

The proportion of YSOs in each class provides a global average estimate of the evolutionary status within a region. Comparing these ratios in different zones can shed light on the history of SF within a region, accounting for the spatial distribution of the objects of different classes.

While the probable members of MYStIX and SFiNCs provide a large, complete, and homogeneously obtained independent sample of YSOs, an evolutionary stage classification of the full sample is impossible, as many members were detected in X-rays and lack IR photometry of sufficient quality. This prevents us from simply counting the elements of each class, and we propose instead a statistical approach for the following analysis. We illustrate this process for region NGC2264 in Figure \ref{relRisk}, composed of 5 panels organised in two rows. In the online complementary materials, described in appendix \ref{appendixCaracts}, we provide additional tests in this region. The large, left panel in Figure \ref{relRisk} shows the spatial distribution of all the sources in the region, coloured according to their class (calculated as described in section \ref{estimEvol}). Green dots stand for Class 0I, red dots for Class II, black dots for Class III, and empty circles for objects classified as non-stellar or unclassified.

First, we apply a fixed-bandwidth Gaussian kernel density estimate (KDE) for the spatial distribution of YSOs. Since the bandwidth choice problem is related to the bias/variance tradeoff problem, a universally optimal parameter does not exist.
As S2D2 computes the relevant small scale of NEST retrieval for each region, $\varepsilon$, we choose the value $h=10\cdot\varepsilon$  to smooth the distribution. This allows us to keep the bandwidth scale related to that of NESTs and consistently use the same approach for all regions.   

Our approach is based on the spatial analysis of a subsample of the members classified as stellar in each region. To obtain robust results, this stellar subsample needs to be representative of the complete population. We flag as unreliable and exclude from further evolutionary analysis all regions where the stellar subsample has less than 50 \% of the members.  We also visually compared the KDE intensity maps from the complete and stellar sample to ensure that both distributions were similar. The top left panel in the right side of figure \ref{relRisk} shows the KDE density estimate of the stellar sample for NGC2264.

 Once the stellar subsample density is deemed representative of the spatial distribution of all members, we use the same bandwidth to compute a non-parametric estimate of relative risk. Relative risk is an exploratory spatial analysis technique widely used in the epidemiological domain \citep[see e.g.][and references therein]{Diggle03} to evaluate the spatial variation of a case of interest such as occurrences of a disease. Relative risk weights the intensity of the case of interest with that of a control (in our case, the stellar subsample). This translates into a quotient of KDEs, providing a continuous and smoothed approach to the ratio of the case of interest within a region. The values of relative risk can be interpreted as the probability of an object at a specific location belonging to the case of interest. In this work, we use as the case of interest the less evolved objects, objects of Class 0/I, but this approach can be applied to any category. 
 
The top right and bottom panels of the right side of Figure \ref{relRisk} show the relative risk maps of Class 0/I (left), Class II (middle) and Class III objects in NGC2264. The relative risk maps expose the distribution of objects, showing a complex SF history for this region    {\citep[consistent with the view provided by numerous previous studies such as e.g][]{Sungetal09,GonzalezAlfaro17, Venuttietal18, Nonyetal21}.}  Class 0/I objects concentrate in the Cone subcluster area towards the South, particularly in the Cone (C) and Spikes areas \citep[where we use the common terminology for the subregions initially proposed by][]{Sungetal09}. Class II show the largest proportions in the Spike subregion, north of Cone-C, and the northern S-Mon subregion. Finally, Class III objects are spread all over, but constitute the majority of sources in the periphery, comprising the Halo and part of S-Mon.
 
 We use R \citep{Rcore23} to apply the relative risk implementations in package \textit{spatstat} \citep[][]{SpatstatPackage, Baddeley15}, including standard edge corrections, and an estimate of the standard error .

\subsubsection{NEST evolutionary estimates and significance}

We estimate the evolutionary state of each NEST from the values of the relative risk maps at its position. We take the position of a NEST as the centroid of the minimum spanning ellipsoid of its members, as \citetalias{Joncouretal18} showed that NESTs are generally small and elongated. The relative risk of  Class 0/I objects in each NEST,  $f_{C0/I}$, is a proxy of its recent activity level and evolutionary stage and consistently traces the SF history within each region.

We note that a straightforward interpretation of the relative risk value suffers from similar issues as ratios obtained from direct counting, where low numbers undermine the reliability of an observed ratio $f_{C0/I}$. We recommend caution, particularly in areas of low YSO density. 
This work focuses on NESTs, which are by definition high density and mitigate this problem, but it should be kept in mind when interpreting the complete relative risk distribution. The consistent application of relative risk maps to different regions in this work will prove that they are a powerful, valuable tool to explore the SF history of a region. 

The standard error estimate map for the relative risk is calculated using the variances of the estimated intensities of the analysed point pattern. While this estimate is valuable, and we provide it for each NEST, it is tied solely to the specific analysed pattern, without controlling for significance w.r.t a spatial random distribution. We follow the philosophy of the S2D2 procedure and determine a global significance level for the  $f_{C0/I}$ values in each region using a random distribution of the class labels.


We redistribute the labels randomly within the stellar subsample and recompute the corresponding relative risk maps using the same parameters. For each resampling we obtain a distribution of relative risk values $r$, with average and standard deviation $\bar{f}, \sigma_{f}$. We perform this procedure 100 times, obtaining 100 relative risk maps $f_i,~ i=1, ... 100$. The mean values of their averages and deviations $\bar{f}_i, \sigma_{f_i},~ i=1, ... 100$ are used as the global average and dispersion control values for each region.
The value $\bar{f}_{C0/I}$ is very close to the ratio of class 0/I within a region $\frac{N_{C0/I}}{N_{class}}$, where ${N_{C0/I}}$ is the number of class 0/I objects and ${N_{class}}$ the number of classified objects: $|\bar{f}_{C0/I}-\frac{N_{C0/I}}{N_{class}}| \sim \mathcal{O}(10^{-3})$. The average dispersion $\sigma_{f_{C0/I}}$ is a conservative estimate of the uncertainty, as it is at least a factor of 2 larger than the mean of the relative risk standard error map in all of the regions in this work.

Given that all sources were extracted from observations with the same instruments, differences in distance are associated with the spatial resolution. We expect an impact on the size of the observed regions, on $\varepsilon$, the scale of NEST retrieval, and on the values of $\bar{f}_{C0/I}$. We explore these biases in the online complementary material described in Appendix \ref{appendixCaracts} and will limit comparisons of ${f}_{C0/I}$ in NESTs to their host region, which is expected to be fair and independent of distance.  {A NEST has a significantly high (resp. low) $f_{C0/I}$ value w.r.t. its host region if $f_{C0/I}> \bar{f}_{C0/I}+\sigma_{f_{C0/I}}$ (resp. $f_{C0/I}<\bar{f}_{C0/I}-\sigma_{f_{C0/I}}$ ).}

\section{Results} \label{results_general}
The combination of MYStIX and SFiNCs has 39 regions, but we have excluded a region with a very low number of members, LDN1251B. In this section, we discuss the results for the resulting sample of 38 regions.
\subsection{General structure and characteristics}

 The main structural characteristics of the regions and spatial analysis results are summarised in table \ref{tableSummaryRegions}. A complete table with all the individual values is available in the appendix \ref{appendixCaracts}. The general characteristics of regions portray a very heterogeneous sample, with values of several physical features spanning more than an order of magnitude. Distances range from 235 to 3600 pc, and the number of objects $N \in [63,~2790]$. $N_{merged}\in [63,~2708]$ was obtained after merging objects within $2 ^{"}$  to account for multiples and chance alignments that could lead to spurious NEST retrieval. We also calculated a typical size $R$ for each region, as the radius of the convex hull of its members, which ranges from 0.61 to 21.1 pc. Given the distance range and its associated resolution limit, we separately study its influence on some key parameters and results on the online supplementary material.

\begin{table}[ht!]

\centering
\caption{Summary statistics of structural characteristics. $d$ is the distance in pc, $R$ is the radius in pc, $N$ is the number of members of each region, $N_{merged}$ is the number of members after merging objects within $2 ^{"}$ , $Q$ is the structural parameter introduced by \citealt{CartwrightWhitworth04}, $N_{NEST}$ is the number of retrieved NESTs, and $\varepsilon$ and $N_{min}$ are the scale and number parameters for DBSCAN determined by S2D2. $\rho_{rel}$ is the relative density parameter, $f_{NEST}$ is the fraction of stars in NESTs within a region, and $N_{MX}$ is the ratio of the maximum number of members of a NEST and $N_{min}$}
\scalebox{0.92}{
\begin{tabular}{lllllll}
  \hline
  & Min&Q$_1$&Q$_2$&Mean&Q$_3$&Max\\
   \hline
   $d$ (pc)&235.0& 436.2 &830.0  & 1058.2 &1460.0  &3600.0 \\
   $R$ (pc)&0.611  & 1.941 & 3.157 & 4.162 & 4.878 & 21.086 \\
   $N$&63.0 &221.8  & 283.5 &573.1  & 940.8 & 2790.0 \\
    $N_{merged}$&63.0  &219.5  & 282.5 &558.0 & 908.0 & 2708.0 \\
  $Q$ & 0.431 & 0.710 &0.835  &0.804  & 0.892 & 1.008 \\
   $N_{NEST}$&  0.00 & 2.25 & 4.50 & 6.68 & 8.75 & 23.00 \\
  $N_{min}$ & 4.0 & 4.0 & 5.0 & 4.5 & 5.0 & 5.0 \\
  $\varepsilon$  (kAU)& 3.447 & 8.199 & 13.089 & 14.855 &18.919  & 39.341 \\
   $\rho_{rel}$& 10.55  &15.32  &16.65  & 17.77 & 18.45 & 37.74 \\
  $f_{NEST}$& 0.000 & 0.075 & 0.108 & 0.107 & 0.139 & 0.304 \\
 $N_{MX}$&0.000  & 1.562 & 2.800 & 4.457 & 5.188 & 26.750 \\
\hline
\end{tabular}
}
\label{tableSummaryRegions}
\end{table}

After applying S2D2, we found 254 NESTs, which are not homogeneously distributed amongst regions. The maximum number of NESTs retrieved in a region was reached in Eagle, with 23 NESTs, while in NGC1333 and ONC-Flank N, we did not retrieve any significant structure.
We verified the compactness of NESTs using the relative density, the ratio between nominal retrieval density of NESTs $\rho_{nom}$ and that of the region $\rho_{CSR}$: $\rho_{rel}=\frac{\rho_{nom}}{\rho_{CSR}}$.  $\rho_{rel}$ is larger than 10 for all regions and reaches 37.74, confirming the compactness of NESTs in all regions. As for the retrieval characteristics of S2D2, $N_{min}$ stays stable with low values between 4 and 5, and the differences between regions reflect on the physical size scale $\varepsilon$, which ranges from 3.45 to 39.34 kAU, with relative nominal density $10.55 \leq \rho_{rel} \leq 37.74$.

 As a first global approach to analysing the structure of each region, we have calculated the $Q$ parameter, introduced by \citet{CartwrightWhitworth04}. The $Q$ parameter is a numerical index to evaluate whether regions display substructure, single concentrations, or are indistinguishable from CSR within a box-fractal/radial model paradigm. Although limited in terms of discerning capability, it can broadly distinguish substructured, clumpy distributions ($Q\lessapprox 0.7$) from large-scale concentrations ($Q\gtrapprox 0.87$) and CSR ($Q\approx 0.8$). We refer the reader to \citealt{Parker18} and to the appendix in \citetalias{Gonzalezetal21} for a more detailed discussion on the power, limitations, and alternatives to the $Q$ parameter.

\begin{figure}[ht!]
\begin{center}
\includegraphics[width=1.05\columnwidth]{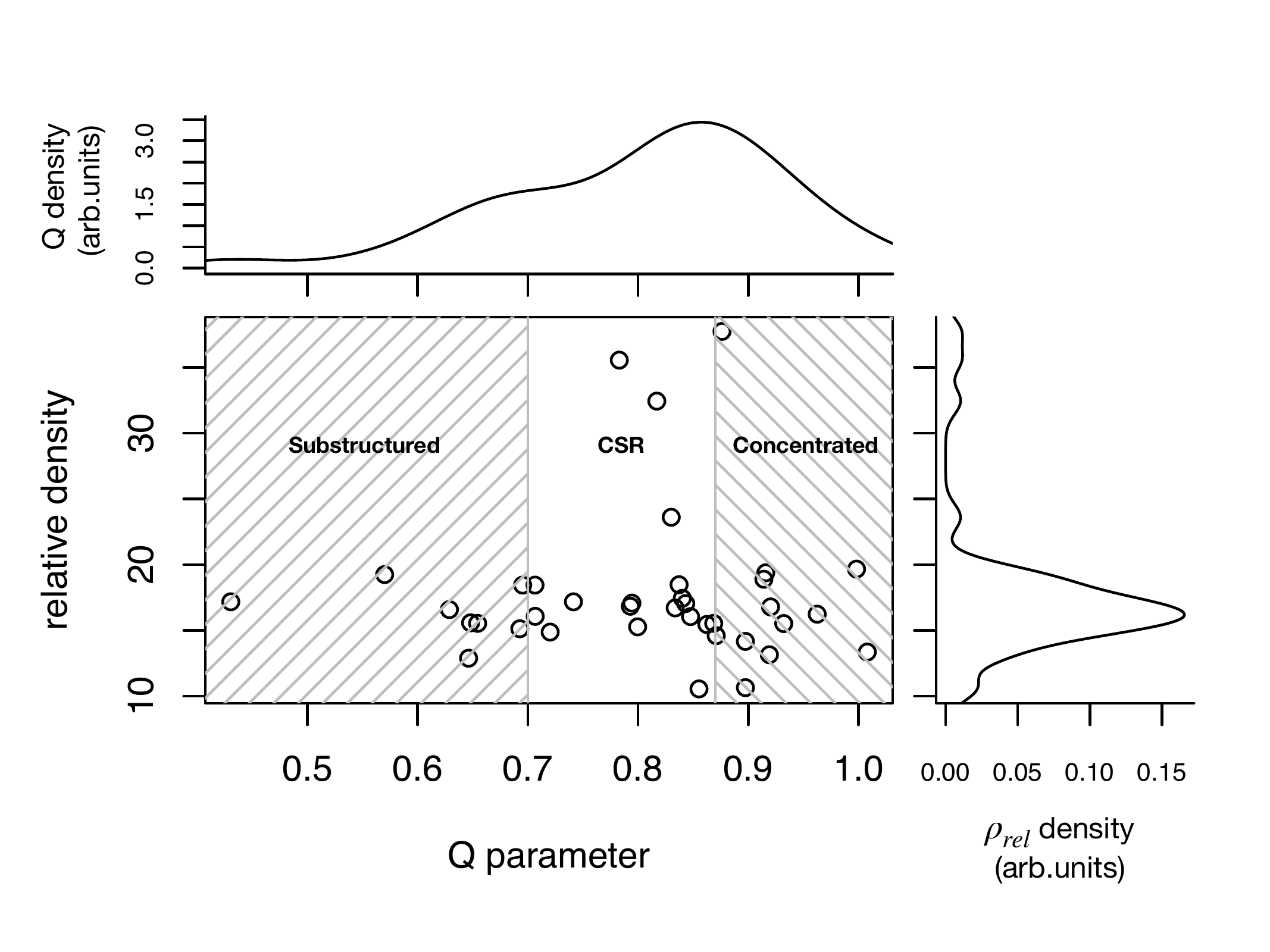}
\end{center}
\caption{Relative density of NESTs compared to that of the region  $\rho_{rel}$  vs. $Q$ structural parameter. The marginal distributions are shown as density profiles on their respective axes.  The hatched zones indicate values of $Q$ corresponding unequivocally to substructured  ($Q\lessapprox 0.7$) and concentrated ($Q\gtrapprox 0.87$)  distributions.}
\label{FigQRho}
\end{figure}

  Figure \ref{FigQRho} shows the relative density values $\rho_{rel}$ against $Q$, with the marginal distributions as density profiles in their respective axes. 
  The global structure traced by the parameter $Q$ indicates significant diversity amongst regions, with $0.43 \leq Q \leq 1.01$. The distribution of $Q$ is spread and peaks at 0.85, displaying a shoulder around 0.65. In Figure \ref{FigQRho} we have hatched the background of the regions with $Q \le 0.7$ (resp. $Q \geq 0.87$), showing clear signs of structure (resp. central concentration) in the box-fractal/radial paradigm. It is noteworthy that those thresholds are close to the quartiles of the $Q$ distribution, meaning that approximately half of the regions are in a range consistent with CSR and low structure levels. As cautioned in \citetalias{Gonzalezetal21}, as $Q$ was calibrated in the box-fractal/radial paradigm, realistic complex regions with substructures can present values of $Q$ consistent with CSR and need to be studied with special care. 

  The distribution of $\rho_{rel}$ is mainly concentrated between 10 and 20, peaking around 15. We note three regions with $\rho_{rel}>30$, namely NGC1333, ONC-Flank N, and SFO2. The high-density requirement along with values of $Q\sim 0.8$ in NGC1333 and ONC-Flank N are consistent with the CSR simulations from \citetalias{Gonzalezetal21}, supporting the validity of the lack of significant structure detection in these regions. SFO2 has $Q=0.88$ pointing to a large-scale concentration traced by a single NEST.

  Other characteristics shown in table \ref{tableSummaryRegions} such as the fraction of stars in NESTs $f_{NEST}$ and $N_{MX}=\frac{N_{max}}{N_{min}}$, which represents the population of the largest NEST in each region relative to ${N_{min}}$, also span ample ranges and their distribution supports the notion that we have a varied sample with significantly different levels of substructure and concentration.

\subsection{Evolutionary stage classification: Regimes of recent activity}\label{sectResEvolGlobal}

We use the ratio of less evolved objects ${\frac{N_{C0/I}}{N_{class}} \sim \bar{f}_{C0/I}}$ as a global indicator of recent SF activity and, as expained in section \ref{methodEvol}, conservatively flagged the classification of all regions where the ratio of classified objects $f_{class}<0.5$. 
Regions flagged for classification are Orion, M17, NGC6334, RCW38, NGC1893, and Trifid. These will be excluded from all discussions regarding recent star formation activity.  

\begin{table}[ht!]
\centering
\caption{Summary statistics of the ratios of objects classified as stellar $f_{class}$ 
and average and standard deviation estimates of the fractions of stellar objects $~\bar{f}_{C0/I},~ \sigma_{f_{C0/I}},~\bar{f}_{CII},~ \sigma_{f_{CII}},~\bar{f}_{CIII},~ \sigma_{f_{II}} $  in the region sample with accepted classifications. }
\begin{tabular}{lllllll}
  \hline
  & Min&Q$_1$&Q$_2$&Mean&Q$_3$&Max\\
   \hline
  $f_{class}$& 0.522 & 0.775 & 0.884 &0.816 &0.916  &0.979\\
  $\bar{f}_{C0/I}$& 0.001 &0.039  & 0.077 &0.096 & 0.135 &0.275\\
  $\sigma_{f_{C0/I}}$& 0.007 &0.029  & 0.043 &0.045 &0.058  &0.124\\
  $\bar{f}_{CII}$& 0.053 &0.442  & 0.516 & 0.477& 0.577 &0.651\\
  $\sigma_{f_{CII}}$& 0.021 & 0.064 & 0.075 &0.082 &0.097  &0.186\\
  $\bar{f}_{CIII}$& 0.205 &0.318  & 0.378 &0.427 & 0.497 &0.934\\
  $\sigma_{f_{CIII}}$& 0.022 & 0.062 &0.072  & 0.078&0.092  &0.157\\
\hline
\end{tabular}
\label{tableSummaryRegionsEvol}
\end{table}

\begin{figure*}[ht!]
\begin{center}
\includegraphics[width=0.9\textwidth]{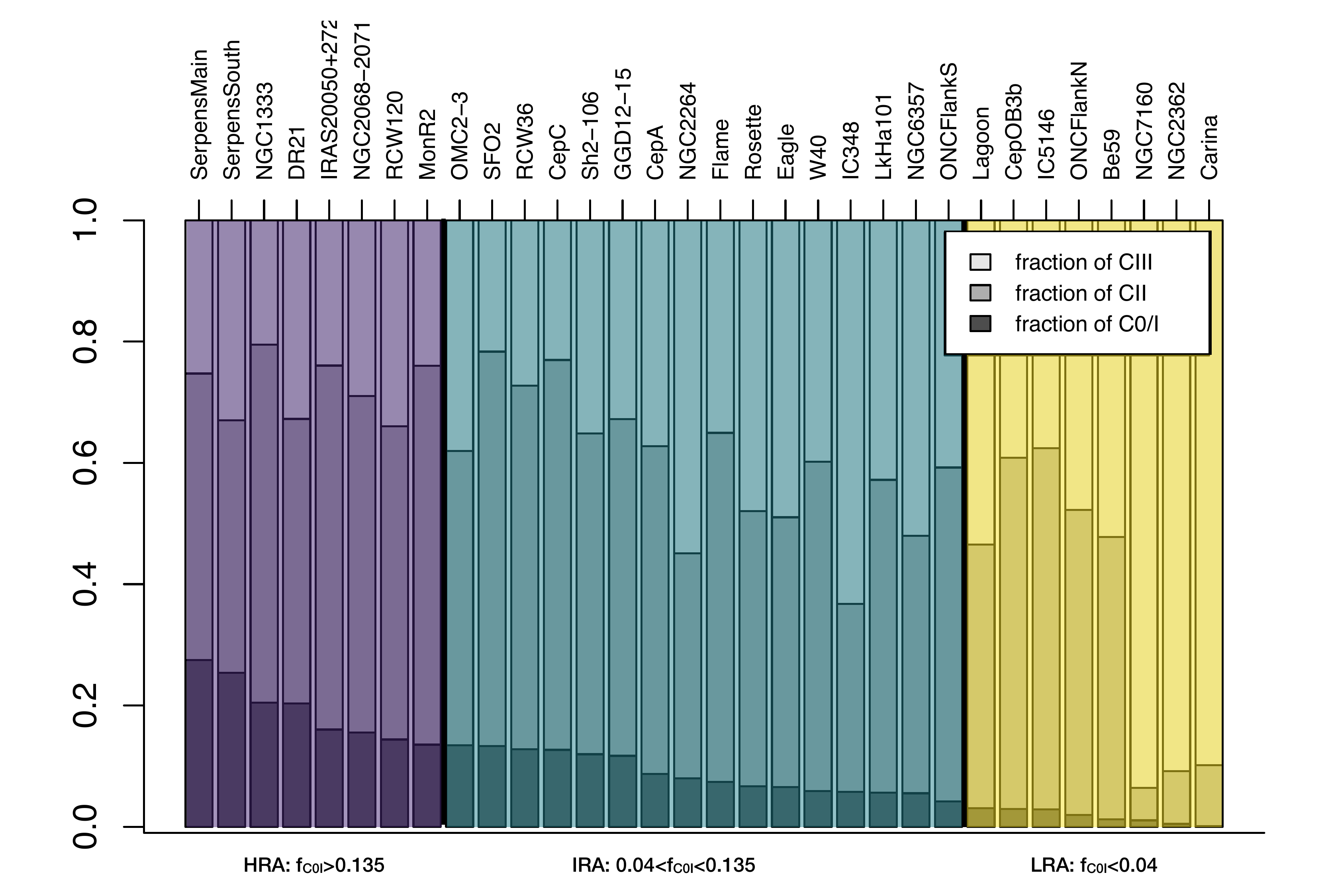}
			
\end{center}
\caption{Stacked bar diagram representing the fraction of stellar objects of each evolutionary state in each (not flagged) region, ordered by Class 0/I fraction. The colour of the bars represents the regime attributed to each region: purple for HRA regions, green for IRA regions and yellow for LRA regions, as explained in the main text. The intensity of each bar represents the different classes of objects in each region, with darker shades representing less evolved objects.}
\label{barPlotAll}
\end{figure*}

The global statistics of the 32 remaining regions involving the evolutionary stage classification, namely $f_{class}$ and the estimates for the relative risk average and standard deviation for each class  $\{  \bar{f}_{A}, \sigma_{f_{A}}\} ,~ A \in \{C0/I,~CII,~CIII\} $,    are summarised in table \ref{tableSummaryRegionsEvol}. 
The values in table \ref{tableSummaryRegionsEvol} show that $C0/I$ objects are the minority in all regions. This can be explained by the difference in average lifetimes for each kind of objects: 0.5 Myr for Class $0/I$ and 2 Myr for Class $II$ \citep[][]{Evansetal09}. With a constant SF rate, the ratio of  $C0/I$ objects $f_{C0/I}$ steadily decreases and drops already below 0.5 after 1 Myr, and 0.25 at 2 Myr. The ratio $f_{CII}$ increases until 2 Myr and decrases from 2.5 Myr onward, as the oldest Class II transition to Class III. After only 4.5 Myr, in this toy model the ratios are $f_{C0/I}=0.1,~ f_{CII}=0.4$, and $f_{CIII}=0.5$.

We separate the regions in the sample into 3 different categories, according to ${\bar{f}_{C0/I}}$. This separation into regimes of recent activity will allow us to simplify our discussion by organizing all the information derived from our different analyses. We considered using age-based criteria but there are large differences between estimates by various authors as well as age spans within regions which prevent an objective, consistent separation \citep[see e.g.][]{Mendigutiaetal22}. Despite the limitations of $f_{C0/I}$, this criterion can be applied homogeneously to all sampled regions.

 \begin{enumerate}
 \item {\textbf{High recent activity (HRA):} 8 regions with $\bar{f}_{C0/I}\ge 0.135$}
 \item {\textbf{Intermediate recent activity (IRA):} 16 regions with $0.04<\bar{f}_{C0/I}\leq 0.135$}
 \item {\textbf{Low recent activity (LRA):} 8 regions with $\bar{f}_{C0/I}\leq 0.04$}
 \end{enumerate}

The chosen boundaries are quartiles of the $\bar{f}_{C0/I}$ distribution of regions with reliable classification, producing a simple and compatible split with qualitative notions of high-intermediate-low activity, but other choices can also be valid. The purpose of defining these regimes is to functionally and broadly categorise regions consistently and facilitate further analyses. 

Figure \ref{barPlotAll} summarises the results of the global evolutionary stage YSO classification for all regions with reliable classification. 
The fractions of objects of different classes are shown as stacked bars, and the regions are ordered in decreasing order according to the value of $\bar{f}_{C0/I}$. The colours represent the recent regime of SF activity in each region, and the shade of each colour represents each class. Darker bars represent $\bar{f}_{C0/I}$, medium bars $\bar{f}_{CII}$, and the lightest colour represents the fraction $\bar{f}_{CIII}$.

\subsection{Evolutionary stage of NESTs} \label{resultsEvol}
In this section we use the proportion of class 0/I objects to estimate the evolutionary stage of NESTs and explore the SF history of each star-forming region.

As described in section \ref{methodEvol}, we statistically estimate the fraction of Class 0/I objects within a NEST, $f_{C0/I}$, and compare it to the average of the region, $\bar{f}_{C0/I}$.  
We then classify NESTs as high-activity, average, or low-activity based on whether $f_{C0/I}$ is significantly larger, consistent, or significantly lower than $\bar{f}_{C0/I}$. 

\begin{table}[ht!]
\caption{Classification of NESTs according to their evolutionary stage estimated from $f_{C0/I}$, grouped by the recent activity regime of its host region.}
\scalebox{0.88}{
\begin{tabular}{ccccc}
Regime & \begin{tabular}{c}
High-activity \\ NESTs
\end{tabular}& \begin{tabular}{c}
Average \\ NESTs
\end{tabular} &\begin{tabular}{c}
Low-activity \\ NESTs
\end{tabular}& Total \\
\hline
HRA    &   34&   8&     2& 44\\    
IRA    &   30&  62&    11& 103\\     
LRA    &   2&  43&   0& 45\\     
\hline
Total & 66& 113& 13& 192\\
\end{tabular}
}
\label{tableEvol}
\end{table}

\begin{figure*}[h!]
\begin{center}                                       
\includegraphics[width=0.8\textwidth]{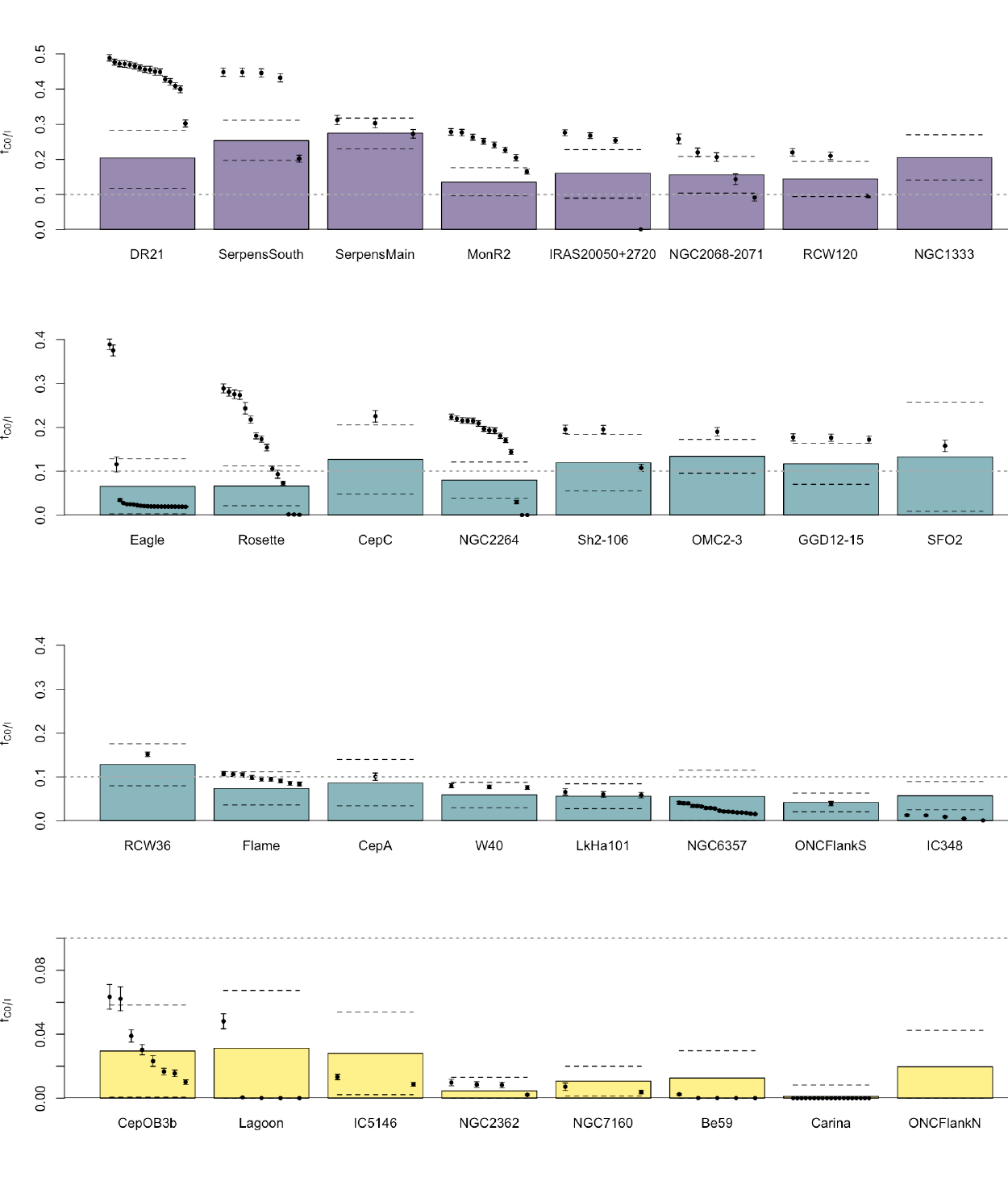}
			
\end{center}
\caption{Bar plots display the average fraction of C0/I objects in each region, with dotted black lines indicating the limits for categorizing NESTs as high-activity or low activity, that is $\bar{f}_{C0/I} \pm \sigma_{{f}_{C0/I}}$. The colour of the bars represents the regime of each region, and black dots represent the values of $f_{C0/I}$ in each NEST, along with its corresponding standard deviation, in decreasing order.  Regions within each regime are ordered by $ f_{NEST, max}$,  the maximum value of $f_{C0/I}$ in NESTs.  The value $f_{C0/I}=0.1$ has been marked in all panels for comparison.}
\label{barPlotNESTs}
\end{figure*}

The results for all NESTs grouped by regime are shown in Table \ref{tableEvol}. Globally, 34\%  of NESTs trace particularly active areas within a region, 59\% show an average level of activity, and only 7\%  of NESTs have significantly lower activity. 
In HRA regions, with global activity ${f_{C0/I}}>13,5\%$, the majority of NESTs show even higher activity: we find 77\% high-activity NESTs, 18\% average and  5\%  (2 out of 44) low-activity.  In IRA regions, 29\% of NESTs are high-activity, 60\% average, and 11\% low-activity. In LRA regions, we only find 5\% of high activity NESTs, while the other 95\% are average.  
A Pearson $\chi ^2$ test of independence provides a p-value of 0.0005, confirming to a high degree of confidence that the distribution of NEST recent activity varies with the SF regime of its host region.

The main results in Table \ref{tableEvol} are that NESTs do not typically trace areas significantly less active than the region average (7 \%), and also that the proportion of high-activity NESTS decreases with the global recent activity level. Table \ref{tableEvol} also points to significant variations in the activity of NESTs within a region. Indeed, the span of $f_{C0/I}$ of NESTs within each region, $\Delta f_{NEST}=\underset{NESTs}{\max}{f_{C0/I}}-\underset{NESTs}{\min}{f_{C0/I}}$, ranges from 0 to 0.37. In 12 regions, we find it to be significant, with $\Delta f_{NEST}>\sigma_{{f}_{C0/I}}$. These account for half of the regions where $\Delta f_{NEST}$ can be calculated (24 regions with reliable evolutionary classification and more than one NEST).

We explore the recent history of SF through the evolutionary stage estimates of NESTs in Figure \ref{barPlotNESTs}, which has 4 panels. Each panel represents regions within each regime, in decreasing order of the maximum value of  $f_{C0/I}$ for  NESTs,  $\underset{NESTs}{\max}{f_{C0/I}}$.  The top panel shows HRA regions, the two middle panels correspond to IRA regions, and the bottom panel shows LRA regions. Each region is represented by a bar with height corresponding to the average value ratio of the region ($\bar{f}_{C0/I}$ ) and coloured according to its regime (purple for HRA, green for IRA, and yellow for LRA).  The significance thresholds $\bar{f}_{C0/I} \pm \sigma_{{f}_{C0/I}}$ in each region are represented by black dashed lines. Black points show the Class 0/I ratio for each NEST ${f}_{C0/I}$, with its standard deviation as error bars. NESTs are ordered decreasingly by their ${f}_{C0/I}$ values. 

Globally, Figure \ref{barPlotNESTs} shows that in most regions, there are NESTs with larger $f_{C0/I}$ than average. Indeed, that is the case for $\sim 76\%$ of regions with NESTs (23 out of 30 regions with reliable classification and NESTs), and in almost half of them (14 out of 30) there are NESTs significantly more active than average. The ratio of more active NESTs also decreases with the global level of activity. 
 
The top panel of Figure \ref{barPlotNESTs} shows the results for all HRA regions with NESTs, ordered by $\underset{NESTs}{\max}{f_{C0/I}}$. In all HRA regions with NESTs, at least some of them have larger $f_{C0/I}$ than average. This difference is significant in most of them, meaning that they host high-activity NESTs. The only exception is Serpens Main, where one NEST is very close to the significance threshold but does not cross it. We only find low-activity NESTs in IRAS20050+2720 and NGC2068-2071,  both regions known as substructured regions hosting 2 clusters, with all the low-activity NESTs grouped in one. 

We find significant values of the span of NESTs $\Delta f_{NEST}>\sigma_{{f}_{C0/I}}$ in 6 out of 7 regions with NESTs. The remaining is again Serpens Main, with $\frac{\Delta f_{NEST}}{\sigma_{{f}_{C0/I}}}=0.89$. Figure \ref{barPlotNESTs} shows both smooth patterns and sudden drops in the $f_{C0/I}$ distribution of NESTs in HRA regions. 
The spatial distribution of NESTS will determine whese patterns can be attributed to either activity gradients, separate episodes of SF, or different subclusters. The general results of such analysis will be described in the next section (and detailed for each region in Appendix \ref{appendixRegions})

The middle panels of Figure \ref{barPlotNESTs} show the results for all 14 IRA regions. Four IRA regions have a significant NEST activity span $\Delta f_{NEST}>\sigma_{fC0/I}$, and 6 regions have a single NEST. Most regions (13/16) have at least one NEST with a larger $f_{C0/I}$ than average, significant in 7 of them. As with HRA regions, we find high-activity NESTs in all regions with $\Delta f_{NEST}>\sigma_{fC0/I}$, accompanied by average and/or low-activity NESTs. In the rest of the IRA regions with more than one NEST, all have similar levels of activity, generally average. The exceptions are IC348, where all the NESTs are low-activity, and GGD12-15, where all NESTs are high-activity.

Finally, the bottom panel from \ref{barPlotNESTs} shows the results for the 8 IRA regions. Here, the global ratio of Class 0/I objects is low, and the dispersion $\sigma_{fC0/I}$ is of the order of the estimate $\bar{f}_{C0/I}$. Only Cep OB3B shows a significant span of activity and hosts NESTs with significantly high activity. The activity of the rest of NESTs in LRA regions is consistent average. 

Our results show NESTs often trace particularly active areas within a region, with higher-activity NESTs being more frequent in regions with high levels of recent activity. 

\section{Spatial distribution of NESTs}\label{results_spatial}

In this section, we show results related to the spatial distribution of NESTs within each region in combination with the evolutionary stage estimates.
Individual detailed results for all regions are shown in Appendix \ref{appendixRegions}, with additional online complementary materials available. 

\subsection{Comparison with  \citetalias{Kuhnetal14} and \citetalias{Getmanetal18}}\label{sectComp}

\begin{figure*}[h!]
\begin{center}
\includegraphics[width=0.85\textwidth]{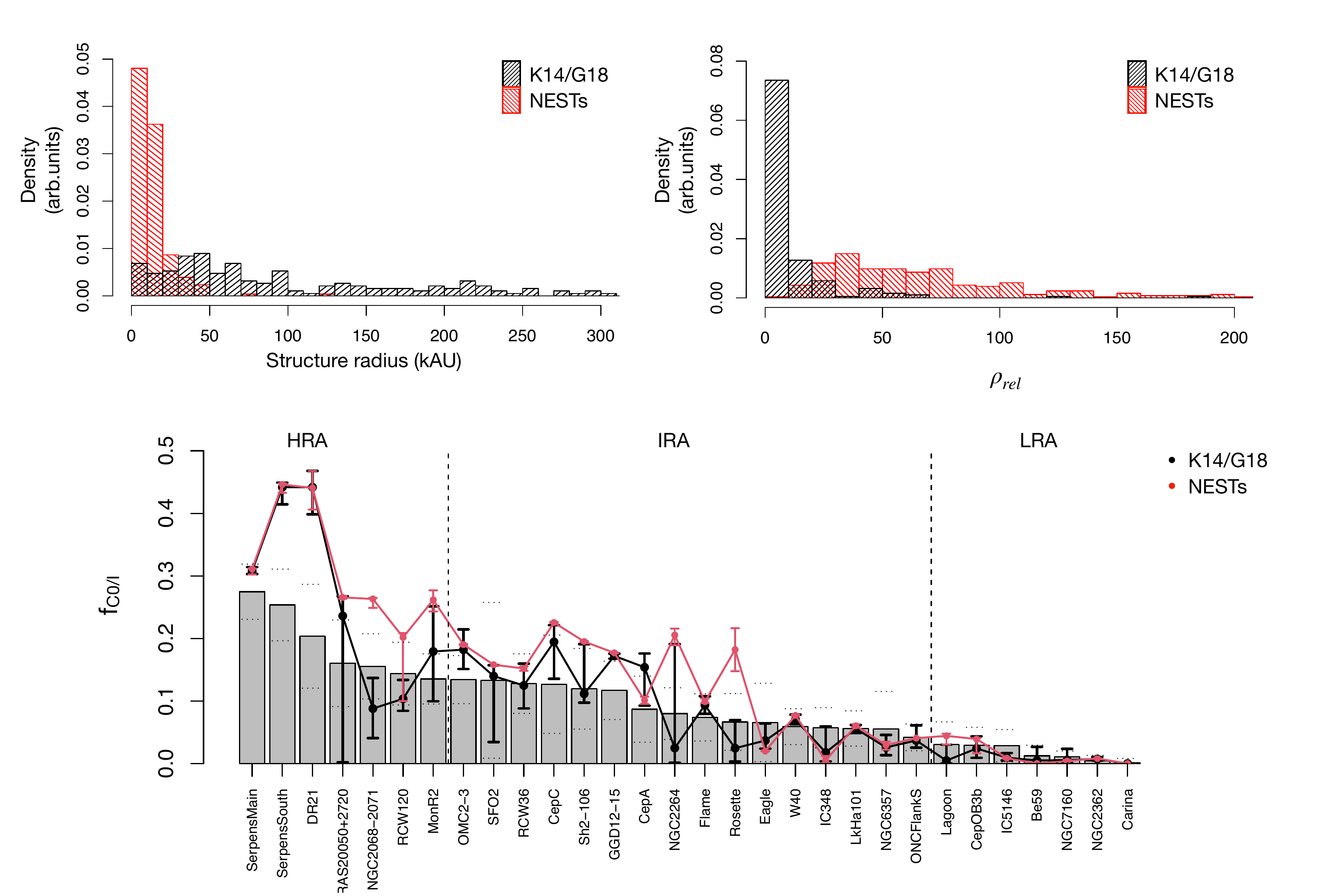}
			
\end{center}
 \caption{ {Comparison of the structures from  \citetalias{Kuhnetal14} and \citetalias{Getmanetal18}. Top left: Histogram of the size of NESTs (red) and the structures from  \citetalias{Kuhnetal14} and \citetalias{Getmanetal18} in kAU.  Top Right: Histogram of relative average density and the structures from  \citetalias{Kuhnetal14} and \citetalias{Getmanetal18}. Bottom: Grey bars show the average ratio  $\bar{f}_{C0/I}$ in each region,  with small dotted horizontal dotted lines showing the limits $\bar{f}_{C0/I} \pm \sigma_{{f}_{C0/I}}$.  Red (resp.  black) dots show the median values of ${f}_{C0/I}$ from the relative risk maps in the area occupied by the convex hull of NESTs (resp. structures from  \citetalias{Kuhnetal14} and \citetalias{Getmanetal18}), with error bars representing the first and third quartile.  Red (resp.  black) solid lines join the median values of ${f}_{C0/I}$ for NESTs (resp.  \citetalias{Kuhnetal14} and \citetalias{Getmanetal18}) to help compare the values.  Finally, vertical lines show separate between HRA, IRA, and LRA regimes. }}
\label{compK14}
\end{figure*}

 {
The natural comparison of our results is with the work \citetalias{Kuhnetal14} and \citetalias{Getmanetal18} that use the same region catalogs. In the supplementary materials, described in Appendix \ref{appendixCaracts},  we also compare with other clustering methods, namely the MST method shown in \citet{Getmanetal18} and HDBSCAN.
\citetalias{Kuhnetal14} and \citetalias{Getmanetal18} fit a mixture model with an unclustered component and isothermal ellipsoids of varying densities and sizes.  The best fit is selected with the Akaike Information Criterion (AIC),  and each ellipsoidal component of the final configuration is considered a substructure. }
 {
The main conclusions of the tests performed in \citetalias{Gonzalezetal21} for Carina  and described in section \ref{s2d2} remain,  even when expanding the sample of regions: NESTs trace the densest parts within each region and are usually close to the core ellipsoids from 
\citetalias{Kuhnetal14} and \citetalias{Getmanetal18}, and NESTs do not detect some of the structures from \citetalias{Kuhnetal14} and \citetalias{Getmanetal18}. } 
 { There are arguments supporting that some of these structures can be spurious as the AIC does not take into account significance,  modeling a single irregular structure may require several ellipsoidal components that do not forcefully correspond to a proper substructure,  and in their solutions there is always at least a cluster.  On the other hand, S2D2 is not sensitive to halos or other real large-scale secondary substructures that do not reach its strict density requirements. \citetalias{Gonzalezetal21} showed that for complex regions where a single comparison density is not appropriate, separating the region before applying S2D2 could lead to the detection of more structures. }

  {Regarding the global number of objects in structures, 65\% of the sample is attributed to a proper substructure in \citetalias{Kuhnetal14} and \citetalias{Getmanetal18} (that is, excluding their field component X and stars with unclear membership),  while only 11\% of all objects are in NESTs.  Only 4\% of stars in NESTs (101 objects) are on the field/unclear component, either part of NESTs that partially overlap with proper structures or small, compact isolated NESTs. We note that the majority of these objects both in NESTs and field/unclear components from \citetalias{Kuhnetal14} and \citetalias{Getmanetal18} are in regions such as M17, Eagle, or RCW38, which have several \citetalias{Kuhnetal14} and \citetalias{Getmanetal18} substructures that overlap in core-halo or multiple clump configurations, and have high ratios of objects with unclear membership. }

  {In Figure \ref{compK14} we compare the general characteristics of the structures from each method.  We use common estimates that can be calculated for any clustering solution and consider the differences between regions.   In the top plot, the histogram compares the sizes of the structures, calculated as the equivalent radius of its reported members in kAU. The distribution of sizes is clearly different, with the NEST distribution very skewed towards small values and the distribution of clusters from \citetalias{Kuhnetal14} and \citetalias{Getmanetal18} much more spread. The situation is reversed for  the relative density distribution histogram of the retrieved clusters,   $\rho_{rel}=\frac{\rho_{str}}{\rho_{CSR}}$, where the density of a structure is given by its number of members and the area of its convex hull,  $\rho_{str}=\frac{N_{str}}{a}$, and $\rho_{CSR}$ is the CSR characteristic density of its host region.  
 From the middle panel of Figure \ref{compK14} it is clear that the distribution of $\rho_{rel}$ of NESTs is much more spread and peaks at higher values than that of clusters from \citetalias{Kuhnetal14} and \citetalias{Getmanetal18}, very skewed towards low values. }

  {The bottom panel of Figure \ref{compK14} compares the distribution of $f_{C0I}$ in NESTs and structures from \citetalias{Kuhnetal14} and \citetalias{Getmanetal18} for all regions with NESTs and reliable classification.  The figure shows the average  $f_{C0I}$ in each region as a gray bar, and horizontal dotted lines mark the limits $\bar{f}_{C0/I} \pm \sigma_{{f}_{C0/I}}$.  On top of the bars,  Figure \ref{compK14}  shows the median value of the $f_{C0/I}$ relative risk map within the convex hull of NESTs (red dots) and clusters \citetalias{Kuhnetal14} and \citetalias{Getmanetal18} (black dots) ,  with an error bar of the corresponding colour spanning the first and third quartiles.  }
  { The aggregated results from Figure \ref{compK14} are consistent with those for individual NESTs shown in Figure \ref{barPlotNESTs},  showing a trend between the activity of NESTs relative to average  and the global level of activity of a region.  Structures from \citetalias{Kuhnetal14} and \citetalias{Getmanetal18} do not show such a clear pattern, and being larger, their distribution of $f_{C0/I}$ is also more spread.  The median $f_{C0I}$ in NESTs is similar to or larger than  $f_{C0I}$ in structures from \citetalias{Kuhnetal14} and \citetalias{Getmanetal18} in all of HRA regions and most IRA and LRA regions,  with the exceptions of Cep A,  Eagle, and IC348, where NESTs have lower median values.  For IC348 and Cepheus A,  figure \ref{barPlotNESTs} already showed values of $f_{C0I}$ consistent with or significantly lower than the region average, and for Eagle, only 2 out of 23 NESTs have significantly high activity,  so their high $f_{C0I}$ values are not expected to appear in the part of the distribution shown.}
 
   {The comparison with \citetalias{Kuhnetal14} and \citetalias{Getmanetal18}  confirms that S2D2 consistently retrieves small scale, very dense substructure characterised by high activity levels, at least for the less evolved regions.  We note again that NESTs leave behind less compact structures,  which could be part of a more complete description of star-forming regions.} 
 
\subsection{HRA regions}

\begin{figure}[ht!]
\begin{center}
\includegraphics[width=0.9\columnwidth]{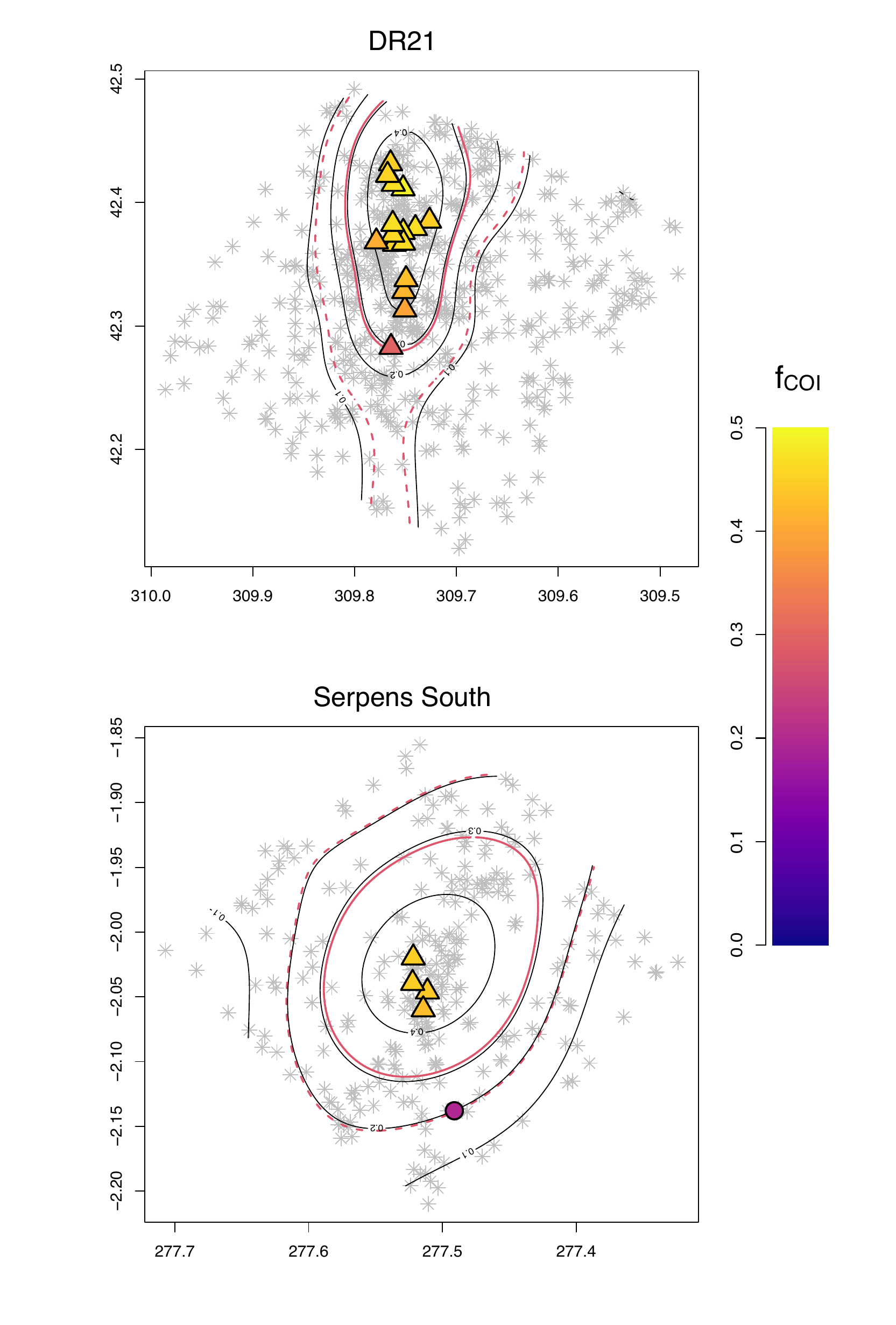}
			
\end{center}
\caption{Maps of selected HRA regions with NESTs. Grey dots represent the YSOs in the area, and coloured markers show the position of NESTs. Symbols indicate whether the significant activity of NESTs is significantly different (triangles for larger, and inverted triangles for lower), or consistent with the average (circles). The colour scale shows the ratio of Class 0/I objects assigned to each NEST. Contours show each region's relative risk map values, and red contours mark the threshold boundaries  $\bar{f}_{C0/I} \pm \sigma_{{f}_{C0/I}}$ separating areas with values different from the average control distribution for each region (solid red contour for the upper limit and dashed for the lower). }
\label{mapsHRA_sel}
\end{figure}
The YSO distribution in most HRA regions shows one or more large-scale concentrations traced by NESTs. There are 6 regions with significant $\Delta f_{NEST}$, which allow us to trace gradients in the SF activity of large-scale structures (Serpens South, DR21, Mon R2), different subclusters(IRAS200050+2720, NGC2068-2071), and feedback effects (RCW120).

The HRA regions with gradients are particularly interesting, as they form elongated patterns which also overlap with dense gas structures. Promising preliminary work \citep{Gonzalezetal21Coolstars} explored the relationships between NESTs and the gas distribution using Herschel \citep{Herschel} data of surveys focusing on star formation \citep[such as HOBYS,][]{Motteetal10Hobys}, and we intend to further investigate it in the future.  
Figure \ref{mapsHRA_sel} shows maps of two HRA regions showing NESTs chains: DR21 and Serpens South.
Each panel displays the YSOs within a region as grey asterisks and the centroids of NESTs as coloured symbols. Circles indicate NESTS of average activity, triangles high-activity NESTs and inverted triangles low activity NESTs. The colour of each NEST corresponds to its value of $f_{C0/I}$ according to the colourbar to the right. Contours show the relative risk map of Class 0/I YSOs. Red contours correspond to the threshold values $\bar{f}_{C0/I} \pm \sigma_{{f}_{C0/I}}$, separating the areas of significantly high (solid line) or low activity (dashed line). We show analogous plots for all regions, along with a detailed analysis in Appendix \ref{appendixRegions} and larger versions of the maps as complementary materials, as described in Appendix \ref{appendixCaracts}.

Both DR21 and Serpens South display a large group of NESTs in the center of the region following an elongated pattern and with significantly high activity ($f_{NEST}\gtrapprox 0.4$), while the more peripheral NESTs have significantly lower $f_{C0I}$ values, indicating an outward decreasing activity ratio. Region DR21 has a richer substructure, displaying several sub-chains along the main concentration.  

\subsection{IRA regions}

\begin{figure}[ht!]
\begin{center}
\includegraphics[width=0.9\columnwidth]{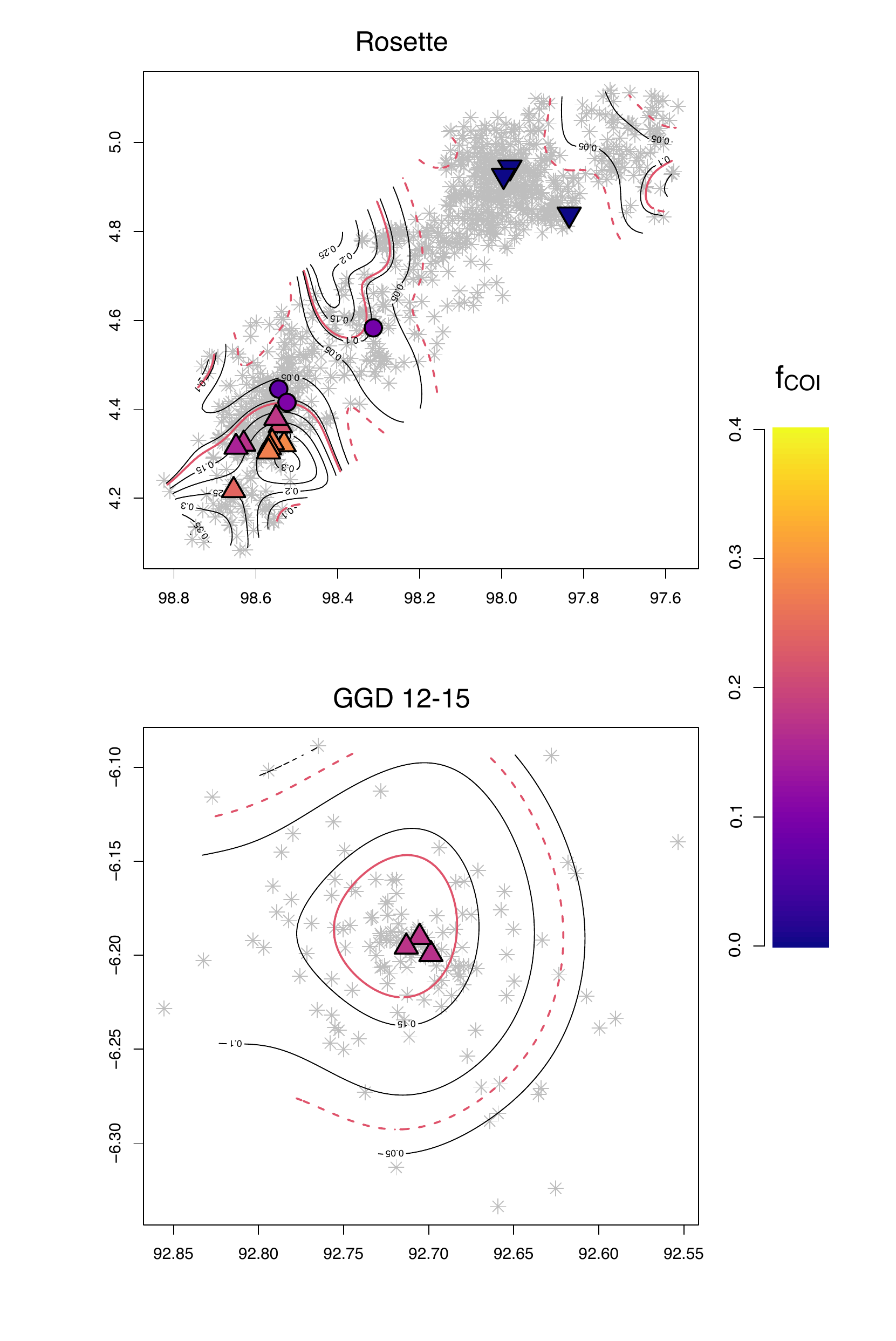}
			
\end{center}
\caption{Maps of selected IRA regions with NESTs, with the same symbols as in Figure \ref{mapsHRA_sel}. }
\label{mapsIRA_sel}
\end{figure}

In IRA regions, we can distinguish two basic patterns in the spatial distributions of YSOs, and consequently NESTs. We show representative examples of both in Figure \ref{mapsIRA_sel} . Each panel shows a map of a region,  displaying the YSOs as grey asterisks and the centroids of NESTs as coloured symbols, with the same code as in Figure \ref{mapsHRA_sel}.

The first group comprises IRA regions showing clear signs of substructure traced by distinct groups of NESTs.  These regions have larger sizes ($R \gtrapprox 4$ pc), most of them also contain high-activity NESTs with high $f_{C0/I}$ values, and their NEST population shows a significant span of activity $\Delta f_{NEST}$.  The top panel of Figure \ref{mapsHRA_sel} shows such a pattern in the Rosette nebula, where we find several separate groups of NESTs whose values of $f_{0/I}$ globally increase towards the South.  The largest population is the southernmost one, which shows a similar configuration to that of HRA regions discussed in the previous section: an elongated configuration of high-activity NESTs with an outward decreasing gradient of recent activity.

The second pattern in IRA regions corresponds to smaller fields ($R \lessapprox 3$ pc) showing a much simpler history.  Their distribution of YSOs is consistent with some level of spatial concentration, and we find a single NEST or group of NESTs, all with similar recent SF activity, usually higher than average (significantly so in approximately half of the cases). An example of this is shown in  the bottom panel of Figure \ref{mapsIRA_sel}.
Only NGC6357, where we trace 3 distinct subclusters from NESTs with similar average activity, does not fit in any of the patterns we just described.

 \subsection{LRA regions}
 
 \begin{figure}[ht!]
\begin{center}
\includegraphics[width=0.87\columnwidth]{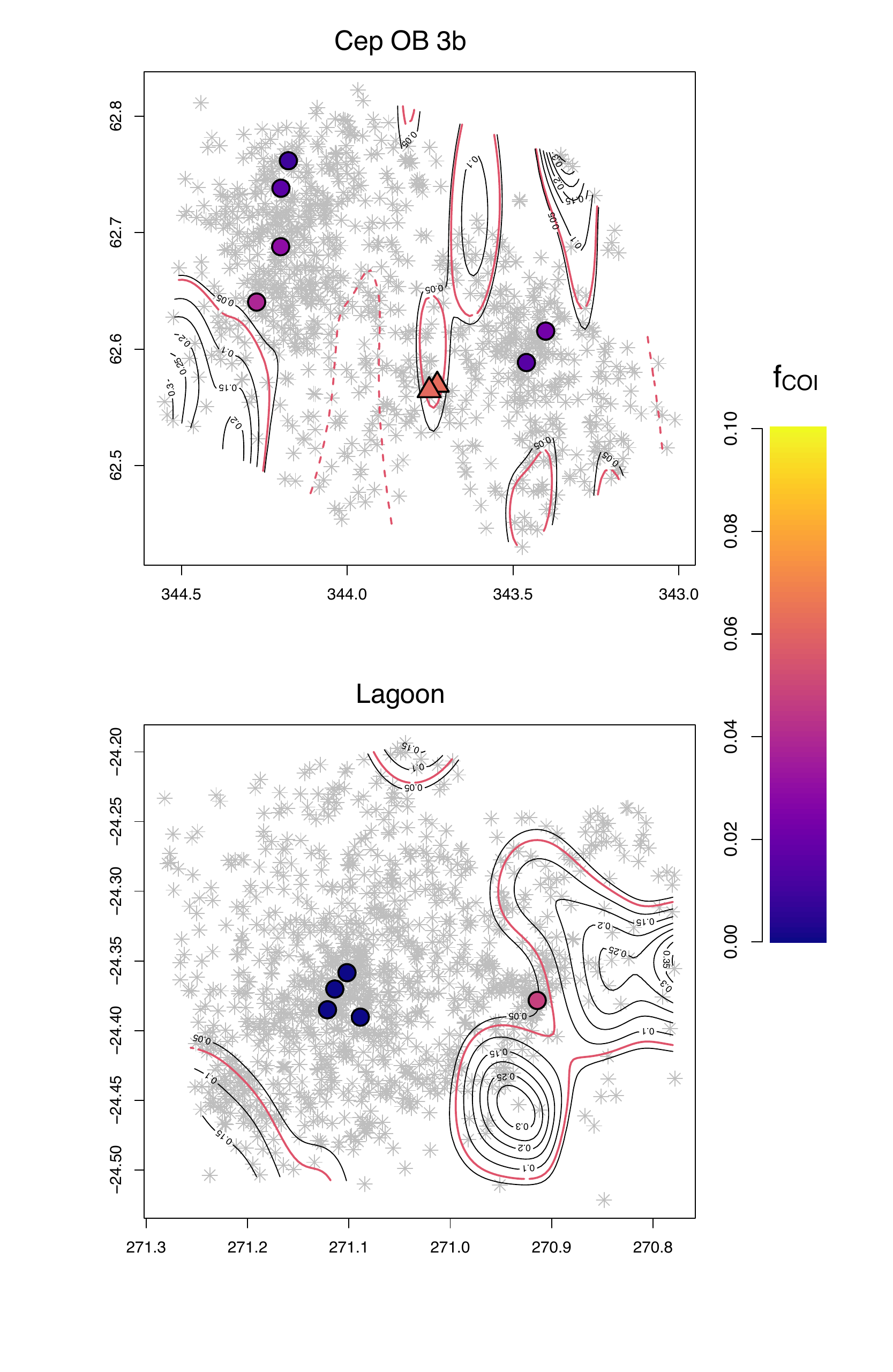}
			
\end{center}
\caption{Maps of selected LRA regions with NESTs, with the same symbols as in Figure \ref{mapsHRA_sel}. }
\label{mapsLRA_sel}
\end{figure}

Structurally, LRA regions are varied. We find spatially distinct groups of NESTs in half of the regions, mainly corresponding to known clusters and large-scale structures. 
Figure \ref{mapsLRA_sel} shows the maps of LRA regions, Cep OB 3b and NGC2362, with the same symbols as previous figures \ref{mapsHRA_sel} and \ref{mapsIRA_sel}.

LRA regions are characterised by their low global ratio of Class 0/I objects, indicating low levels of recent star formation, and the dispersion $\sigma_{{f}_{C0/I}}$ is of the order of $\bar{f}_{C0/I}$. For these regions individually, the ratio of Class 0/I + Class II objects could be a more informative indicator, but in this work, we keep the $f_{C0/I}$ for the sake of general consistency in the comparison with other regions.

The recent SF history we trace in LRA regions is simple: we only find NESTs with activity levels different from average in 1 region out of 8, Cep OB3b. In this region, shown in the top panel \ref{mapsLRA_sel}, we find a pair of high-activity NESTs at the centre of the field, while NESTs towards the east and west have average activity. In LRA regions, recent star formation plays a lesser role, and the retrieved significant structure can be associated with older SF events that dominate the YSO distribution. An example of such a situation is NGC2362, shown in the bottom panel of Figure \ref{mapsLRA_sel}.

\section{Discussion}\label{discussion}

In this section, we discuss the role of NESTs in the study of SF, focusing on their interpretation as remnants of the star formation process. 

Choices in the implementation of S2D2 were oriented towards robustness in structure retrieval (see the online complementary material described in Appendix \ref{appendixCaracts} for a detailed comparison with other methods of structure extraction), but as any spatial technique applied in 2D, it can be affected by projection effects. 
Projection effects are ultimately a lack of information that can be extremely relevant, depending on the unknown underlying situation. Specific studies show that the trends detected by INDICATE are robust to both evolutionary and projection effects \citep{Buckneretal24, Buckneretal22, BlaylockSquibbsetal24}.  In \citetalias{Gonzalezetal21} we verified that S2D2 is consistent with INDICATE, and the retrieved NESTs are all in areas of strong clustering tendencies. We note, however, that specific further analyses (which are out of the scope of this work) are needed to verify that the most clustered stars, as those that would be in NESTs, are preserved in projection. 

\subsection{NESTs as imprints of star formation}

Both observations \citep[e.g.][and references therein]{Sanchezetal24, Buckneretal20} and simulations \citep[e.g.][and references therein]{Pelkonenetal24, Verliatetal22} support a higher level of substructure for the spatial distribution of less evolved YSOs. Younger stars are typically grouped towards the densest gas, while the older ones show a more homogeneous distribution. The procedure S2D2, by construction, recovers the denser substructures that we expect from younger objects.

The potential of NESTs as tracers of recent SF is also substantiated by star formation scenarios where small structures of young objects appear as a consequence of fragmentation. A recent semi-analytical model for hierarchical fragmentation by \citet{Thomassonetal24} predicts structures compatible with the size and number of objects in NESTs found in Taurus by \citetalias{Joncouretal18} as well as with the cloud fragments retrieved in  NGC2264 \citep{Thomassonetal22}.   {The size distribution of NESTs (top left panel of  Figure \ref{compK14}) is very skewed towards low sizes,  and has median radius ~10 kAU,  which is compatible with fragmentation imprints.  As \citetalias{Joncouretal18}  proposed for Taurus,  the smaller NESTs  could be the product of the fragmentation of a single molecular core,  while the larger ones may trace clusters of cores.}

 {The work from \citet{VazquezSemadenietal25} compares the current state of the Global Hierarchical Collapse (GHC) model with turbulent support (TS) scenarios. Both are hierarchical models comprising turbulence and gravity, but the roles and effects of these processes are fundamentally different. The main difference between both views is the overall state of the molecular cloud. In the GHC there is a global contraction and fragmentation, while in the TS model turbulence produces an almost hydrostatic cloud. While there are not specific predictions regarding small scales, the GHC expects a gravitational fragmentation cascade which can get sizes consistent with NESTs \citep{VazquezSemadenietal19}. } 
 {  We also lack information at smaller scales for the TS, but the related turbulent core SF model from \citet{McKeeTan03} for massive star formation, where fragmentation of cores has low theoretical probabilities, primordial NESTs are not expected.}

  {We remind the reader that, as described in section \ref{sectResEvolGlobal}, we use estimates of the ratio $f_{C0/I}$ to determine of the level of recent activity for both NESTs and regions and that we separate regions in high, intermediate, and low recent activity (HRA, IRA and LRA) according to their global ratio of class 0/I objects.}
Our general results show that while it is not uncommon to find highly active NESTs ($\gtrapprox 30\%$), low-activity NESTs are rare as they constitute a $7\%$ of the total. Moreover, the evolutionary distribution of NESTs within a region is dependent on its global regime of activity: the less globally evolved a region, the greater the chances of finding high-activity NESTs within it. In HRA regions, most NESTs are high-activity, and most HRA regions contain at least one high-activity NEST. Both the overall ratios of high-activity NESTs and regions containing them decline with the global level of activity.

If we consider NESTs as potential pristine remnants of SF, their persistence is key, as the general picture emerging from both observations and simulations for the early evolution of young clusters is very dynamic.  While we lack specific analysis at small scales, simulations show that structures often appear,  interact,  dissolve,  merge, and split while new stars still form during the embedded phases \citep[see e.g.][]{Guszejnovetal22, Dobbsetal22}.  After a few Myrs, when the cloud starts dispersing, structures typically expand, although structure remains can persist.  Associations seem particularly stable, retaining significant levels of substructure despite their expansion \citep{CantatGaudin24}. 

\citet{MiretRoigetal24} found a discrepancy of $\sim 5$ Myr between the isochronal and dynamical ages in 6 young clusters, attributed to the embedded phase delaying the onset of pure dynamical evolution. These timescales are larger than the average expected lifetimes of Class 0/I and II objects, supporting limited effects of dynamical evolution in our samples, at least for those without large Class III ratios.

In active regions like DR21,  MonR2,  or Serpens, we found elongated structures of high-activity NESTs overlapping with dense gas. This spatial configuration is similar to some found in IRA regions (southern group of NESTs in Rosette) or without a reliable classification (e.g.  NGC1893 or NGC6334) and in Taurus by \citetalias{Joncouretal18}.  In Taurus,  NESTs were also aligned with the gas elongated structures,  as would be expected for the remnants of the fragmentation of filaments. 
Large groups of NESTs overlapping hubs or ridges of dense gas (like NGC2264,  Orion,  M17, Flame, or GDD 12-15)  are also common.  

The spatial distribution of NESTs matches the evolutionary morphological path proposed by \citetalias{Kuhnetal14}, in particular the predominance of string-like structures of NESTs associated with gas in highly active regions.  \citetalias{Kuhnetal14} proposed that linear subcluster chains inherited directly their structure from gas filaments and evolved towards more concentrated morphologies, such as those that would be traced by a NEST (or a group of them). The spatial distribution of NESTs is also compatible with the high-mass formation scenario outlined in \citep{Motteetal18Review}, where ridges and hubs undergo a global collapse which increases the masses of both the ridge/hub and its forming protostars.

In scenarios where large structures of YSOs form by the assembly of smaller substructures \citep[as suggested by simulations by e.g.][]{Dobbsetal22, VazquezSemadenietal17},  NESTs could either detect the large-scale structures themselves or some lasting remains of the individual components, depending on the level of dynamical mixing. Analysis of kinematical data could help constrain the dynamical status of NESTs and their members.

Our results support that NESTs, at least those highly active in HRA regions, are the pristine imprints of SF as suggested by \citetalias{Joncouretal18}.  The interpretation of low activity NESTs is less straightforward, as it depends on the level of interactions within each region.  Some can be infertile remnants of earlier star formation, while others could have assembled later on.

\subsection{Tracing SF History within a region}

We have shown that relative risk maps provide a powerful, visual approach to the global SF history within a region. The interpretation of maps should be done carefully, considering the specifics of each sample. In this work, we apply this technique massively and focus mainly on NESTs, particularly compact ones, to maximise the underlying density support. We believe carefully exploring the complete maps can be worthwhile, particularly for structures traced by single NESTs (e.g. Eagle, Sh2-106, Cep A, Cep C, SFO2, RCW36).  

Not all structures traced by NESTs are high-activity, nor associated with relative risk peaks or dense gas substructures. Indeed, we find groups of low-activity (IRAS20050+2720, NGC2078-2071, NGC2264, Rosette), or average NESTs (Eagle, Rosette, W40) even in HRA and IRA regions. These groups could be remnants of earlier SF depleting the local gas reservoirs. However, being regions generally evolved, they may also have suffered a relevant amount of dynamical evolution, and their interpretation needs to consider all the specific aspects in each region. 

Large groups of NESTs may also display significant differences in activity, allowing us to trace evolutionary SF patterns. We find outward gradients continuously traced by NESTs, often associated with highly active, globally collapsing clouds (e.g. DR21, Mon R2, and Serpens South), where large clusters are potentially assembling. 
Similar outward gradients have been observed in several young clusters \citep[see the recent summary by][and references therein]{Stahler24}. \citet{Getmanetal18AgeGradients} also found gradients in substructures retrieved from the MYStIX-SFiNCs samples, and our work supports their findings. 

While such evolutionary gradients can be due to projection effects of sequential SF \citep{Maaskantetal11}, they can also form naturally in some star formation scenarios. In the GHC, they are associated with accretion at all scales: the older stars formed in the filaments are displaced and end up distributed over larger areas than their younger counterparts, forming inside the clump \citep{VazquezSemadenietal17}. 
 {In a TS paradigm, age gradients only form in simulations when gradual SF is included \citep{Fariasetal19}.}
 \citet{Stahler24} proposes outwards diffusion of stars forming in the center due to overcrowding  {that creates a halo-like structure called mantle}. The gradient appears as new stars form in the densest part and different structures, such as open clusters, would be produced by the disruption of the mantles. 
 In the slingshot mechanism \citep{StutzGould16}, the magnetic field produces a transverse wave oscillating along a filament with active SF. Newborn objects follow the oscillations and inherit the transverse velocity as they decouple from the gas. 
 {Some of these scenarios overlap the 8 options that \citet{Getmanetal14}, grouped in three blocks. The first one involves more recent star-formation in the center of the structure, due to the gas density gradient, acceleration of the SF rate with time, or an age stratification like the one reported for high mass stars. The second option involves older stars moving outward and basically includes scenarios related to the dynamical evolution of the YSOs. These include drift or relaxation from supervirial or subvirial initial configurations, and the effects of initial substructure as older clusters expand more. In the final alternative, younger stars move inward, either as infalling filaments or subcluster mergers. While we cannot discard the presence of other effects, the appearance of groups of NESTs tracing large scale substructures alligns our results with the infalling fillament and subcluster merger scenarios.}

We also find large-scale evolutionary patterns consistent with sequential SF in regions such as NGC2264, Rosette, or with major feedback effects, as in RCW120. In approximately 1/3 of the regions, we find spatially distinct groups of NESTs often coincident with known clusters. This is the case for larger regions  {($R \geq 4 $ pc)},  where NESTs outline the most significant overdensities.   
These structures generally have significantly different levels of recent activity, principally in HRA and LRA regions,  making NESTs effective tracers of different SF episodes within each region.  
The variety and complexity of the patterns that we find are supported by the spatio-kinematical analysis from \textit{Gaia} data reporting the appearance of large-scale spatio-temporal patterns in associations \citep[such as e.g.][]{Kerretal24, Ratzenbocketal23}. Their results point to both the continuous formation of low mass clusters and episodes of heightened SF with several Myr separations. 

 {This work, which includes just spatial and evolutionary analysis of YSO candidates, can solely provide a partial view on SF, a very complex and widespread phenomenon that involves several processes extremely hard to quantify. We expect such an event to adopt different forms depending on the specific conditions, so studies including kinematics as well as characteristics of the maternal cloud are required to obtain a more complete picture. Despite these limitations, our results favour a hyerarchical scenario comprising fragmentation such as the GHC. This would be compatible with the structure we find at all spatial scales and the scenarios in \citet{Getmanetal14} explicitly including large scale substructures, such as infalling filaments or subcluster mergers.}

\section{Summary and conclusions}\label{conclusions}

{We create a catalogue of NESTs, significant small-scale substructures, retrieved by S2D2 in a homogeneous sample of YSOs of 38 star-forming regions. 
The regions vary in size, distance, population, activity, and structure. The properties of the NESTs retrieved within them are also heterogeneous. We find a total of 254 NESTs in 36 regions, and none in the remaining 2 (NGC1333 and ONCFlankN), which show other indicators of consistency with CSR.
We do not find significant global patterns of the number of NESTs, Q parameter, fraction of objects within NESTS or relative maximum population of NESTs with the regime of recent SF. }%
 
{We estimate the evolutionary stage of the YSOs in each region using IR photometry and separate the regions with reliable classification in three recent activity regimes depending on their observed ratio of Class 0/I objects, $f_{C0/I}$. We calculate the relative risk of Class 0/I objects (which can be interpreted as the probability of finding an object of Class 0/I at a given position), and propose a statistical indicator of the evolutionary state of the NESTs within each region. Based on this, we determine whether the level of recent activity in each NEST is significantly higher, lower or consistent with the average of the region.}

{Our results support the idea that NESTs can trace the preferential sites of SF in regions with high recent star formation activity. There, NESTs tend to be high-activity, overlap dense gas structures, and are often arranged in elongated patterns or groups consistent with pristine remnants of the fragmentation of filaments and cluster assembly process.The probability of retrieving high-activity NESTs decreases with the global level of recent SF, and the probability of dynamical processing of the traced spatial structures increases.}

{The spatial distribution of NESTs themselves provides useful insights into the most significant large-scale structures. In approximately half of the regions, we find a significant span on the level of recent activity of NESTs. Often this is due to distinct spatial structures that display different activity levels, effectively tracing separate clusters and episodes of SF. We are also able to detect activity gradients inside groups of NESTs that could be attributed to sequential SF, hierarchical collapse, stellar mantle drifts, or feedback effects. }

Our work balances the power of relative risk maps and the high significance of NESTs to trace the remnants of star formation and discern different patterns of SF history.
 {The variety of spatial structures and evolutionary patterns we obtain favours hyerarchical multiscale models comprising fragmentation, such as the GHC. We note, however, that our analysis are limited to the spatial and evolutionary studies, so these conclusions must be supported by additional evidence, notably kinematics and the behaviour of the gas.}
{ Preliminary work shows promising results \citep{Gonzalezetal21Coolstars} on the systematic separation and analysis of groups of NESTs in combination with the study of gas within a region. }

\begin{acknowledgements}
We thank the anonymous referee, whose comments and suggestions improved this manuscript.
This work acknowledges funding from European Union’s Horizon 2020 research and innovation program under grant agreement no. 687528 (STARFORMMAPPER) and the European Research Council (ERC) via the ERC Synergy Grant ECOGAL (grant 855130). This project has received funding from the Spanish Agencia Estatal de Investigación (MICIU/AEI
/10.13039/501100011033) and FEDER, EU through project TACOS ( PID2023-146635NA-I00) as well as the French Agence Nationale de la Recherche (ANR) through the project COSMHIC (ANR-20-CE31-0009).
\end{acknowledgements}
\bibliographystyle{aa} 
\bibliography{bibliographySM}

\begin{appendix}
\section{Useful lists and tables}\label{appendixCaracts}
\subsection{Acronyms}
We start by describing the acronyms used in this work, in alphabetical order 
\begin{itemize}
    \item \textbf{AIC}: Akaike information criterion. Criterion used in \citetalias{Kuhnetal14} and \citetalias{Getmanetal18} to select the number of clusters. It selects a model based on likelihood  penalised by the model complexity. 
    \item \textbf{CSR}: Complete Spatial Randomness. Term from spatial statistics refering to a homogeneous spatial distribution, characterised by constant density. 
        \item \textbf{GHC}: Global Hierarchical Collapse, a star formation scenario proposed by \citet{VazquezSemadenietal19}
    \item \textbf{HRA}: High Recent Activity. Refers to regions displaying signs of significant recent star formation, characterised by high ratios of Class 0/I sources. 
    \item \textbf{IR}: Infrared.
    \item \textbf{IRA}: Intermediate Recent Activity. Refers to regions displaying signs of moderate recent star formation, characterised by intermediate ratios of Class 0/I sources.
    \item \textbf{LRA}: Low Recent Activity. Refers to regions with almost no signs recent star formation, characterised by low ratios of Class 0/I sources.
    \item \textbf{NEST}: Acronym of Nested Elementary Structures, introduced in \citetalias{Joncouretal18}. It refers to compact, small and significant substructures extracted by S2D2 within a star-forming region. 
   \item \textbf{OPCF}: One-Point Correlation Function. Statistical function introduced by \citet{Joncouretal17}. The OPCF is a function of the spatial scale that compares the nearest neighbour distance distribution of a sample with that of a control distribution, typically CSR. 
    \item \textbf{SED}: Spectral Energy Distribution.
    \item \textbf{SF}: Star Formation.
    \item \textbf{S2D2}: Small, Significant DBSCAN Detection, introduced in \citetalias{Gonzalezetal21}. It is a procedure of retrieval of dense small, significant structures from the spatial distribution of stellar members of a region.
     {
    \item \textbf{TS}: Turbulent Support star formation scenario, as described in \citet{VazquezSemadenietal25}}
   \item \textbf{YSO}: Young Stellar Object. 
\end{itemize}

\subsection{Available online supplementary materials}
Tables  \ref{tableAppendix1} and \ref{tableAppendix2} are available at CDS, along with the NESTs catalogue that we describe below. We list and describe the additional materials available online, in \textit{Zenodo} at \url{https://doi.org/10.5281/zenodo.18955037}
\begin{itemize}
    \item     {\textbf{Tables}
        \begin{itemize}
        \item Catalogue of NESTs: This table is available both at CDS and the online repository. Column Region contains the name of the host region, column regime contains the regime of recent activity, column meanf\_C0I\_region is the average value of class 0/I objects in the region $\bar{f}_{C0/I}$, sigmaf\_C0I\_region is the standard deviation of ${f}_{C0/I}$ value in the region, $\sigma_{{f}_{C0/I}}$. Column id\_Nest is the NEST identifier, N\_star is the number of stars in the NEST, RA and DEC are the J2000 coordinates of the NEST centroid, f\_C0I (resp.  f\_CII and f\_CIII) is the  ${f}_{C0/I}$ (resp. ${f}_{CII}$ and ${f}_{CIII}$) estimate in the NEST, and sigmaf\_C0I (resp. sigmaf\_CII and sigmaf\_CIII) is the standard deviation value in the NEST, $\sigma_{{f}_{C0/I}}$
        \item Members of each region. These are available only on the online repository. Each CSV file contains Tables with data of each source in the original catalogues. Column PCM contains the probable complex member identifier in the original MYStIX and SFiNCs catalogues, columns RAJ2000 and DEJ2000 contain its coordinates,column NEST contains the identifier of its host NEST (0 if it does not belong to a NEST), and Class contains its evolutionary classification (Class0I, ClassII, ClassIII for objects classified as stellar, notStars for objects classified as contaminants, and NA for objects that could not be classified).
        \end{itemize}}

    \item \textbf{Additional robustness analysis.} We provide a document with three additional analyses concerning the robustness of our results: 
        \begin{itemize}
            \item Relationships between distance, spatial scales and activity regimes.
            \item Tests on NGC2264 on the evolutionary stage procedure
            \item Comparison with different retrieval methods
        \end{itemize}

    \item \textbf{PDF Maps}
    \begin{itemize}
        \item Spatial distribution of NESTs for each region
        \item Comparison of NESTs with other retrieval methods
    \end{itemize}     
    
\end{itemize}

\subsection{tables}
\begin{sidewaystable*}[hp!]
\begin{tabular}{lrrrrrrrrrrrrr}
  \hline
 Region & RA & DEC & $d$ & $N$ & $N_{merg}$ & $R$ (pc) & $Q$ & $N_{NEST}$ & $N_{Min}$ & $\varepsilon$(kAU) & $\rho_{Rel}$ & $f_{NEST}$ & $N_{MX}$\\%
  \hline
 Be59 & 0.564 & 67.433 &  900 &  433 &  420 & 3.848 & 0.855 &    5 &    5 & 16.004 & 10.550 & 0.110 & 3.400 \\%
  Carina & 161.079 & -59.889 & 2300 & 2790 & 2708 & 21.086 & 0.629 &   19 &    5 & 26.854 & 16.585 & 0.095 & 25.600 \\
 CepA & 344.059 & 62.038 &  700 &  163 &  161 & 2.993 & 0.843 &    1 &    4 & 13.577 & 17.048 & 0.043 & 1.750\\
 CepC & 346.456 & 62.508 &  700 &  131 &  130 & 2.993 & 0.830 &    1 &    4 & 10.926 & 23.606 & 0.046 & 1.500\\
 CepOB3b & 343.914 & 62.636 &  700 & 1019 & 1015 & 4.661 & 0.646 &    8 &    5 & 11.420 & 12.887 & 0.056 & 2.200\\
 DR21 & 309.757 & 42.327 & 1500 &  662 &  660 & 5.018 & 0.800 &   16 &    4 & 16.222 & 15.273 & 0.186 & 4.250\\
 Eagle & 274.700 & -13.807 & 1750 & 1614 & 1573 & 8.179 & 0.793 &   23 &    5 & 18.904 & 16.838 & 0.132 & 4.200\\ 
 Flame & 85.428 & -1.912 &  414 &  342 &  334 & 1.045 & 0.962 &    9 &    4 & 5.572 & 16.236 & 0.153 & 3.000 \\
 GGD12-15 & 92.706 & -6.196 &  830 &  147 &  147 & 1.857 & 0.998 &    3 &    4 & 12.184 & 19.662 & 0.150 & 2.250 \\
 IC348 & 56.126 & 32.129 &  300 &  224 &  224 & 1.246 & 0.834 &    5 &    4 & 7.463 & 16.712 & 0.107 & 1.500 \\
 IC5146 & 328.374 & 47.256 &  800 &  231 &  229 & 3.322 & 0.862 &    2 &    5 & 13.801 & 15.452 & 0.157 & 6.400 \\
 IRAS20050+2720 & 301.777 & 27.501 &  700 &  280 &  280 & 1.967 & 0.720 &    4 &    5 & 10.060 & 14.874 & 0.125 & 3.600\\
 Lagoon & 270.904 & -24.387 & 1300 & 1251 & 1202 & 4.788 & 0.794 &    5 &    5 & 12.980 & 17.079 & 0.046 & 6.000\\
 LkHa101 & 67.541 & 35.269 &  510 &  149 &  145 & 1.284 & 0.837 &    3 &    4 & 9.649 & 18.478 & 0.097 & 1.250\\
 M17 & 275.196 & -16.172 & 2000 & 1322 & 1226 & 4.908 & 1.008 &   19 &    5 & 14.601 & 13.356 & 0.179 & 7.000\\ 
 MonR2 & 91.942 & -6.382 &  830 &  279 &  279 & 2.158 & 0.869 &    8 &    4 & 13.199 & 15.538 & 0.140 & 1.750\\
 NGC1333 & 52.270 & 31.336 &  235 &  118 &  118 & 0.611 & 0.817 &    0 &    4 & 3.848 & 32.440 & 0.000 & 0.000\\
 NGC1893 & 80.683 & 33.412 & 3600 &  853 &  827 & 9.698 & 0.919 &   14 &    5 & 34.049 & 13.150 & 0.116 & 3.200 \\ 
 NGC2068-2071 & 86.681 & 0.126 &  414 &  234 &  233 & 1.959 & 0.570 &    5 &    4 & 10.811 & 19.248 & 0.116 & 2.500\\ 
 NGC2264 & 100.242 & 9.890 &  914 &  969 &  935 & 3.957 & 0.648 &   15 &    4 & 12.011 & 15.578 & 0.137 & 5.500\\ 
 NGC2362 & 109.671 & -24.955 & 1480 &  246 &  244 & 4.001 & 0.932 &    4 &    4 & 23.698 & 15.522 & 0.078 & 1.500\\
 NGC6334 & 260.212 & -36.115 & 1700 &  987 &  980 & 7.436 & 0.706 &    7 &    5 & 19.727 & 18.446 & 0.056 & 3.400 \\ 
 NGC6357 & 261.625 & -34.200 & 1700 & 1439 & 1399 & 8.321 & 0.654 &   18 &    5 & 18.924 & 15.538 & 0.171 & 9.200\\ 
 NGC7160 & 328.458 & 62.586 &  870 &   96 &   96 & 4.117 & 0.897 &    2 &    5 & 33.461 & 10.648 & 0.115 & 1.200\\ 
 OMC2-3 & 83.848 & -5.118 &  414 &  239 &  238 & 1.959 & 0.695 &    1 &    5 & 7.888 & 18.446 & 0.021 & 1.000\\ 
 ONCFlankN & 83.810 & -4.830 &  414 &  219 &  218 & 1.959 & 0.783 &    0 &    4 & 5.463 & 35.559 & 0.000 & 0.000\\ 
 ONCFlankS & 83.774 & -5.652 &  414 &  237 &  236 & 1.959 & 0.707 &    1 &    4 & 7.946 & 16.071 & 0.017 & 1.000 \\
 Orion & 83.822 & -5.391 &  414 & 1367 & 1345 & 1.145 & 0.871 &   13 &    5 & 3.447 & 14.602 & 0.088 & 5.800\\ 
RCW120 & 258.097 & -38.479 & 1350 &  278 &  275 & 4.350 & 0.840 &    3 &    5 & 24.109 & 17.444 & 0.084 & 2.200 \\
 RCW36 & 134.754 & -43.736 &  700 &  306 &  294 & 1.935 & 0.897 &    1 &    5 & 5.329 & 14.165 & 0.122 & 7.200 \\
 RCW38 & 134.773 & -47.511 & 1700 &  495 &  461 & 5.612 & 0.916 &    6 &    4 & 17.276 & 19.385 & 0.304 & 26.750 \\ 
 Rosette & 97.917 & 4.963 & 1330 & 1195 & 1176 & 9.980 & 0.430 &   15 &    5 & 24.821 & 17.178 & 0.082 & 2.600\\ 
 SerpensMain & 277.485 & 1.220 &  415 &  105 &  104 & 0.989 & 0.920 &    3 &    4 & 8.958 & 16.788 & 0.192 & 2.500 \\
 SerpensSouth & 277.511 & -2.062 &  415 &  287 &  285 & 1.122 & 0.741 &    5 &    5 & 7.196 & 17.176 & 0.116 & 2.200 \\
 SFO2 & 1.016 & 68.559 &  900 &   63 &   63 & 2.900 & 0.876 &    1 &    4 & 14.956 & 37.743 & 0.079 & 1.250\\ 
 Sh2-106 & 306.851 & 37.382 & 1400 &  221 &  215 & 4.071 & 0.914 &    3 &    4 & 20.309 & 18.892 & 0.186 & 7.500\\ 
 Trifid & 270.675 & -22.972 & 2700 &  357 &  353 & 7.390 & 0.693 &    3 &    5 & 39.341 & 15.126 & 0.088 & 3.400 \\
W40 & 277.860 & -2.073 &  500 &  411 &  408 & 1.339 & 0.848 &    3 &    5 & 7.497 & 16.050 & 0.074 & 3.800 \\
   \hline
\end{tabular}
\caption{Characteristics of regions. Column region: Name of each region.  Columns RA and DEC: J2000 coordinates of each field in decimal degrees. Column $d$: Distance from \citetalias{Kuhnetal14} and \citetalias{Getmanetal18}. Column $N$: number of stars in the member flattened sample from \citetalias{Kuhnetal14} and \citetalias{Getmanetal18}. Column $N_{merg}$: Number of stars after merging objects with distance less than $2 ^{"}$ . Column $R$: equivalent radius of the convex hull of members in pc. Column $Q$: structural parameter $Q$ from \citet{CartwrightWhitworth04}. Column $N_{NEST}$: Number of retrieved NESTs. Column $N_{Min}$: NEST retrieval minimum number of points parameter of DBSCAN. Column $\varepsilon$: NEST retrieval scale parameter of DBSCAN in kAU. Column $\rho_{Rel}$: Ratio of nominal DBSCAN retrieval density and characteristic density of the region used for control w.r.t. CSR. Column $f_{NEST}$: fraction of stars in NESTs. Column $N_{MX}$: population of the largest NEST in each region relative to $N_{Min}$.  } 
\label{tableAppendix1}
\end{sidewaystable*}

\begin{table*}[h!]
\begin{tabular}{llrrrrrrrrr}
  \hline
 region & regime & $N_{class}$ & $f_{class}$ & $\Delta f_{NEST}$ & $\bar{f}_{C0/I}$ & $\sigma_{f_{C0/I}}$& $\bar{f}_{CII}$ &  $\sigma_{f_{CII}}$ & $\bar{f}_{CIII}$ &  $\sigma_{f_{CIII}}$ \\ 
  \hline
 Be59 &  LRA &  395 & 0.912 & 0.002 & 0.012 & 0.017 & 0.465 & 0.083 & 0.522 & 0.083 \\ 
  Carina &  LRA & 1740 & 0.624 & 0.000 & 0.001 & 0.007 & 0.101 & 0.100 & 0.898 & 0.101 \\ 
 CepA & IRA &  151 & 0.926 & 0.000 & 0.087 & 0.053 & 0.540 & 0.096 & 0.373 & 0.091 \\ 
 CepC & IRA &  116 & 0.885 & 0.000 & 0.127 & 0.079 & 0.643 & 0.120 & 0.230 & 0.102 \\  
 CepOB3b &  LRA &  951 & 0.933 & 0.053 & 0.029 & 0.029 & 0.579 & 0.084 & 0.392 & 0.083 \\ 
 DR21 &  HRA &  384 & 0.580 & 0.187 & 0.204 & 0.083 & 0.469 & 0.100 & 0.327 & 0.094 \\ 
 Eagle &  IRA &  864 & 0.535 & 0.370 & 0.066 & 0.063 & 0.445 & 0.131 & 0.490 & 0.133 \\ 
 Flame &  IRA &  271 & 0.792 & 0.023 & 0.074 & 0.038 & 0.576 & 0.072 & 0.350 & 0.069 \\ 
 GGD12-15 &  IRA &  137 & 0.932 & 0.005 & 0.117 & 0.047 & 0.555 & 0.072 & 0.328 & 0.070 \\ 
 IC348 &  IRA &  209 & 0.933 & 0.012 & 0.057 & 0.032 & 0.310 & 0.068 & 0.633 & 0.070 \\ 
 IC5146 &  LRA &  214 & 0.926 & 0.005 & 0.029 & 0.026 & 0.596 & 0.076 & 0.376 & 0.074 \\ 
 IRAS20050+2720 & HRA &  246 & 0.879 & 0.275 & 0.160 & 0.069 & 0.600 & 0.095 & 0.239 & 0.084 \\ 
 Lagoon &  LRA &  704 & 0.563 & 0.048 & 0.031 & 0.036 & 0.435 & 0.112 & 0.535 & 0.113 \\ 
 LkHa101 &  IRA &  124 & 0.832 & 0.007 & 0.056 & 0.028 & 0.516 & 0.059 & 0.428 & 0.059 \\ 
 M17 & FLAG &  187 & 0.141 & -- & -- & -- & -- & -- & -- & -- \\ 
 MonR2 & HRA &  250 & 0.896 & 0.114 & 0.135 & 0.040 & 0.625 & 0.061 & 0.240 & 0.050 \\ 
 NGC1333 & HRA &  107 & 0.907 & 0.000 & 0.205 & 0.065 & 0.590 & 0.077 & 0.205 & 0.062 \\ 
 NGC1893 &FLAG &  388 & 0.455 & -- & -- & -- & -- & -- & -- & -- \\ 
 NGC2068-2071 &  HRA &  218 & 0.932 & 0.167 & 0.155 & 0.052 & 0.555 & 0.070 & 0.290 & 0.065 \\ 
 NGC2264 &IRA &  774 & 0.799 & 0.224 & 0.080 & 0.041 & 0.371 & 0.074 & 0.549 & 0.076 \\ 
 NGC2362 &  LRA &  217 & 0.882 & 0.008 & 0.005 & 0.008 & 0.087 & 0.039 & 0.908 & 0.040 \\ 
 NGC6334 &FLAG &  376 & 0.381 & -- & -- & -- & -- & -- & -- & -- \\ 
 NGC6357 &  IRA &  751 & 0.522 & 0.025 & 0.055 & 0.060 & 0.425 & 0.137 & 0.520 & 0.138 \\ 
 NGC7160 & LRA &   94 & 0.979 & 0.003 & 0.011 & 0.009 & 0.053 & 0.021 & 0.936 & 0.022 \\ 
 OMC2-3 &  IRA &  216 & 0.904 & 0.000 & 0.134 & 0.038 & 0.485 & 0.053 & 0.381 & 0.051 \\ 
 ONCFlankN &  LRA &  203 & 0.927 & 0.000 & 0.020 & 0.023 & 0.503 & 0.079 & 0.478 & 0.079 \\ 
 ONCFlankS &  IRA &  213 & 0.899 & 0.000 & 0.042 & 0.021 & 0.550 & 0.048 & 0.408 & 0.048 \\ 
 Orion &  FLAG &  606 & 0.443 &-- & -- & -- & -- & -- & -- & -- \\ 
RCW120 &  HRA &  250 & 0.899 & 0.124 & 0.144 & 0.050 & 0.516 & 0.072 & 0.340 & 0.067 \\ 
 RCW36 & IRA &  187 & 0.611 & 0.000 & 0.128 & 0.048 & 0.599 & 0.071 & 0.273 & 0.066 \\ 
 RCW38 & FLAG &  203 & 0.410 &-- & -- & -- & -- & -- & -- & -- \\ 
 Rosette & IRA &  993 & 0.831 & 0.288 & 0.067 & 0.045 & 0.454 & 0.090 & 0.479 & 0.092 \\ 
 SerpensMain &  HRA &   95 & 0.905 & 0.039 & 0.275 & 0.044 & 0.473 & 0.050 & 0.253 & 0.044 \\ 
 SerpensSouth &  HRA &  173 & 0.603 & 0.246 & 0.254 & 0.057 & 0.417 & 0.065 & 0.330 & 0.062 \\ 
 SFO2 &  IRA &   52 & 0.825 & 0.000 & 0.133 & 0.124 & 0.651 & 0.186 & 0.217 & 0.157 \\ 
 Sh2-106 & IRA &  174 & 0.787 & 0.088 & 0.120 & 0.064 & 0.529 & 0.104 & 0.351 & 0.100 \\ 
 Trifid & FLAG &  152 & 0.426 & -- & -- & -- & -- & -- & -- & -- \\ 
W40 &  IRA &  304 & 0.740 & 0.004 & 0.059 & 0.029 & 0.543 & 0.063 & 0.398 & 0.061 \\ 
   \hline
\end{tabular}
\caption{Evolutionary statistics of regions. Column region: Name of the region. Column regime: Recent activity regime of each region, or FLAG in case the stellar subsample was not representative. Column $N_{class}$: Number of objects classified as stellar. Column $f_{class}$: fraction of objects with stellar classification. Column $\Delta f_{NEST}$: span of estimated ratio of class $0/I$ objects $f_{C0I}$ in NESTs. Column $\bar{f}_{C0/I}$: Average of the estimated fraction of Class $0/I$ objects in the region, calculated as explained in the main text.  $\sigma_{{f}_{C0/I}}$: Standard deviation of the estimated fraction of Class $0/I$ objects in the region, calculated as explained in the main text. Column $\bar{f}_{CII}$: Average of the estimated fraction of Class $II$ objects in the region, calculated as explained in the main text.  $\sigma_{{f}_{CII}}$: Standard deviation of the estimated fraction of Class $II$ objects in the region, calculated as explained in the main text.Column $\bar{f}_{CIII}$: Average of the estimated fraction of Class $III$ objects in the region, calculated as explained in the main text.  $\sigma_{{f}_{CIII}}$: Standard deviation of the estimated fraction of Class $III$ objects in the region, calculated as explained in the main text} 
\label{tableAppendix2}
\end{table*}

\section{Spatial distribution of NESTs by region}\label{appendixRegions}

\subsection{HRA regions}
\begin{figure*}[ht!]
\begin{center}
\includegraphics[width=0.9\textwidth]{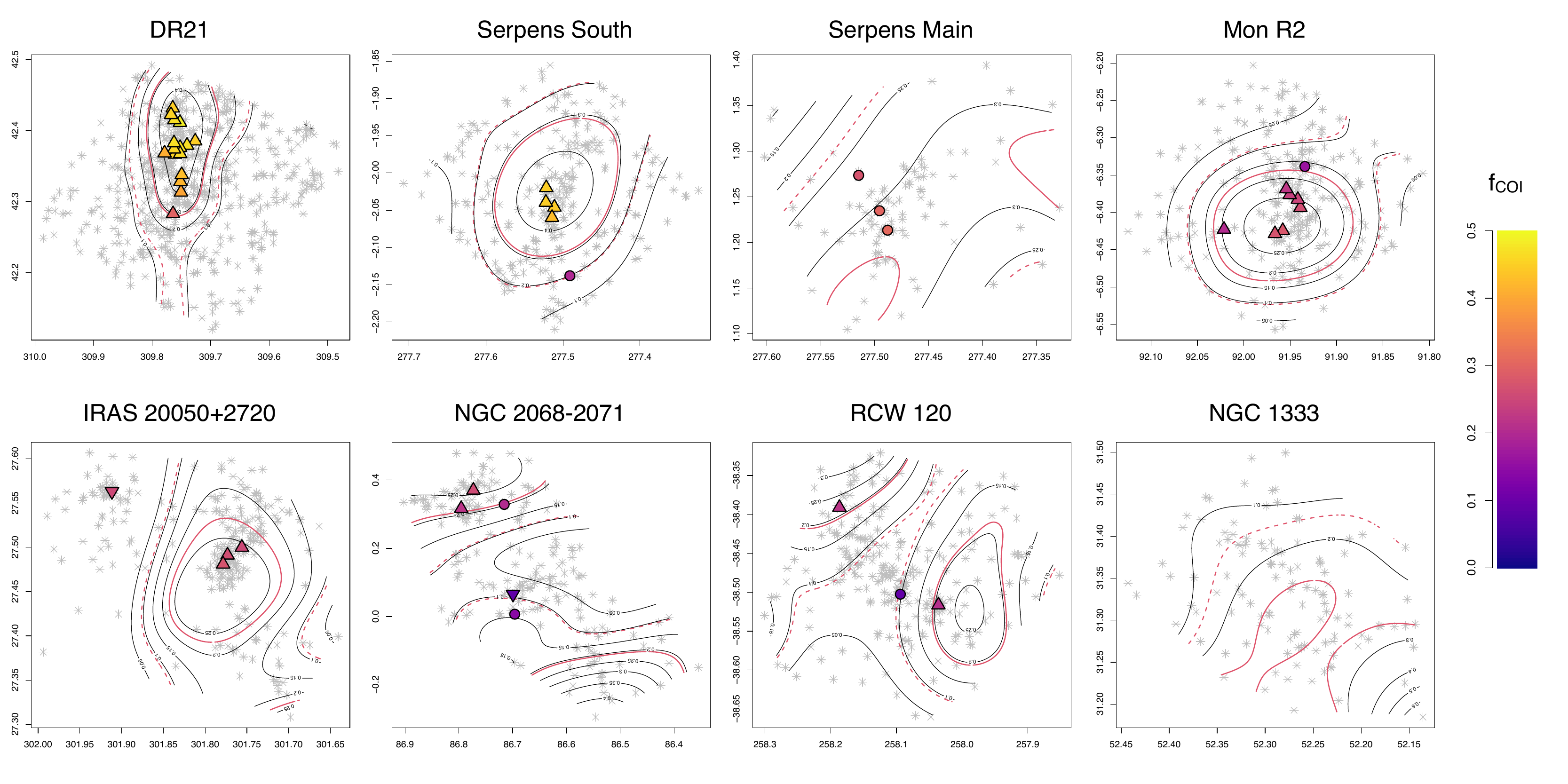}
			
\end{center}
\caption{Maps of HRA regions with NESTs. Grey asterisks represent the YSOs in the area, and coloured markers show the position of NESTs. Symbols indicate whether the significant activity of NESTs is significantly different (triangles for larger, and inverted triangles for lower), or consistent with the average (circles). The colour scale shows the ratio of Class 0/I objects assigned to each NEST. Contours show each region's relative risk map values, and red contours mark the threshold boundaries  $\bar{f}_{C0/I} \pm \sigma_{{f}_{C0/I}}$ separating areas with values different from the average control distribution for each region (solid red contour for the upper limit and dashed for the lower). }
\label{mapsHRA}
\end{figure*}

\begin{enumerate}
    \item{\textbf{DR21}: This is an active protocluster associated with a molecular cloud within the Cygnus X complex. The region is characterised by a ridge and dense filamentary structures \citep{Bonneetal23}. We find a group of 16 NESTs, significantly more active than average, tracing the main concentration of YSOs. The NESTs are arranged in elongated structures and roughly coincide with the location of the denser gas and molecular cores in the ridge \citep{Hennemannetal12}, although we note that robustly assessing this requires further analysis out of the scope of this work.  The relative risk $f_{C0/I}$ peaks in the central area traced by the NESTs and decreases towards the outskirts, and the NESTs trace this outward gradient.  While still significantly more active than average, the southernmost NEST has significantly lower activity than the rest. Despite showing a value $Q=0.8$, the single large-scale structure scenario in this region is supported by the large fraction of $18.6$\% YSOs in NESTs, and the relatively large population $N_{MX}=4.25$. Studies of the morphology and kinematics of the gas component point to a global collapse scenario in the region \citep[][]{Schneideretal10},  which is also consistent with the formation of large-scale concentrations of young objects. }%
    
    \item{\textbf{Serpens South}: The sample is located in a dark cloud filament within the Serpens-Aquila Rift. We find 5 small NESTs hosting 12\% of all the YSOs in the region and showing a significant activity span. There are four highly active NESTs grouped, forming an elongated structure towards the centre of the region that coincides spatially with the  central peak of the relative risk distribution $f_{C0/I}$. It also overlaps with the largest cluster found by the MST and \citet{Getmanetal18}, and with structures of the highest hierarchical level found by \citep{Sunetal22}, which performed a multiscale analysis of a large field in the region, including both Serpens South and W40. The single NEST detected towards the South traces a very compact small structure of average activity close to the low activity significance threshold, also retrieved by \citetalias{Getmanetal18}, and traces the decreasing gradient of the relative risk towards the exterior of the field.  This distant NEST can explain the low value of $Q=0.74$, which is typically associated with some level of substructure.}%
   
    \item{\textbf{Mon R2} We study the main protocluster in the larger Monoceros R2 cloud, within the hub-filament system. The value of Q=0.87 is consistent with a large-scale concentration traced by 8 small NESTs. Most of them are highly active, except for the Northernmost, which is of average activity. NESTs group at the central part of the field, characterised by high values of $f_{C0/I}$, and trace the decreasing gradient towards the periphery.  The distribution of NESTs forms strings that seem to align with the gas and dust substructures \citep{Trevinoetal19,Gutermuthetal11}.  Our results in this region are similar to those in DR21, with a large, high-activity central concentration of NESTs, and suggest global collapse scenarios based on cloud analysis \citep[][]{Trevinoetal19}.}
    \item{\textbf{Serpens Main}: This region has long been known for the concentration and high ratio of Class 0/I objects \citep[see e.g.][and references therein]{Eiroaetal08}, consistent with its classification in the HRA regime. We find a value of $Q=0.92$  supporting a significant central concentration, which is traced by two NESTs of average activity. We also retrieve an additional peripheral NEST towards the North, with average activity. The Northern NEST is small and compact and does not correspond with \citetalias{Getmanetal18} clusters or the MST-retrieved substructures. The global relative risk map is mostly homogeneous: we only find significantly high or low $f_{C0/I}$ values in peripheral areas, but we note that these are close to the borders and have notably low densities. The sample overlaps with part of the Serpens field analysed in \citet{Gutermuthetal11}, and our results are consistent with their cluster, highly correlated with the dust density. }
 \item{\textbf{IRAS200050+2070} This is an active region between the Cygnus Rift and Cygnus X, composed of 2 clusters, named E and W by \citet{Guntheretal12}. We find four NESTs split into two separate groups spatially consistent with the substructures from \citet{Guntheretal12} and \citetalias{Getmanetal18}.  The YSO density is dominated by W cluster, with the value of $Q=0.72$ supporting some level of substructure.  E cluster from \citeauthor{Guntheretal12} is traced by a single NEST with significantly low activity, while three significantly active NESTs trace cluster W and form a string. NESTs trace the main features of the relative risk map, the central high $f_{C0/I}$ area related to the southern part of cluster W, and the low $f_{C0/I}$ towards the North East. Our work is generally consistent with that of \citet{Guntheretal12}, as their sample also contains a large ratio of Class 0/I objects ($\sim 20 \%$), and they do not find Class 0/I objects in subcluster E. Despite that, their isochrone analysis points to a younger age of subcluster E relative to W,  {which shows an age spread.}}
    \item{\textbf{NGC2068-2071} In this portion of the molecular cloud Orion-B, we find two clear groups of NESTs hosting 12\% of the YSOs and corresponding to the two known clusters in the region \citep[see e.g.][and references therein]{Kirketal16}. This configuration is consistent with the low value of $Q=0.57$. The northernmost group, located in NGC2071, has two active NESTs and one of average activity, close to the border. The second group of two NESTs corresponds to NGC2068 and is composed of an evolved and an average NEST. The relative risk map has high $f_{C0/I}$ values in NGC2071, but the central overdensity has significanlty low $f_{C0/I}$ and increases towards the southern boundary. }
    \item{\textbf{RCW 120}  We retrieved 3 small NESTs that only account for 8\% of the stars in the region. The value $Q=0.84$ could support a moderate level of central concentration, but also a situation more complex than the calibration simulations box-fractal/radial model can capture. 
    The North-Eastern and western NESTs are highly active, while the central one is average, although close to the low $f_{C0/I}$ boundary. 
   The global pattern of relative risk is completely decoupled from the density of YSOs, with significantly low values of $f_{C0/I}$ occupying the center of the field and large $f_{C0/I}$ areas towards the North and the West. Complex patterns like this can be explained by environmental feedback effects. Indeed, this region is known as an example of triggered SF, produced by the expansion of an HII region around a central massive star \citep[e.g.][]{Luisietal21} located North of the central, most evolved NEST.}   
    \item{\textbf{NGC1333}: This is a nearby massive protocluster located in Perseus and characterised by a ridge \citep{Hacaretal17}.  We do not find any NESTs within this region. This result, along with the high values of $\rho_{nom}$, and the value $Q=0.82$, points to a spatial distribution consistent with CSR and its random fluctuations. While \citet{Gutermuthetal11} find a clear concentration of objects in his analysis of the Perseus Complex, our analysis is limited to a small part of their field, which can prevent us from tracing the lower density peripheral area of the cluster. 
    Our global ratio $f_{C0/I}=0.21$ agrees with the knowledge of the region, although it is lower than the ratio found by \citet{Young2015}, 0.27. The relative risk map is not homogeneous, with low $f_{C0/I}$ values towards the North and high-risk areas in the center and south. We note, however, that the southern high-risk area has very low density values as support, so these results should be taken with caution. }
\end{enumerate}

\subsection{IRA regions}

\begin{figure*}[ht!]
\begin{center}
    \includegraphics[width=0.9\textwidth]{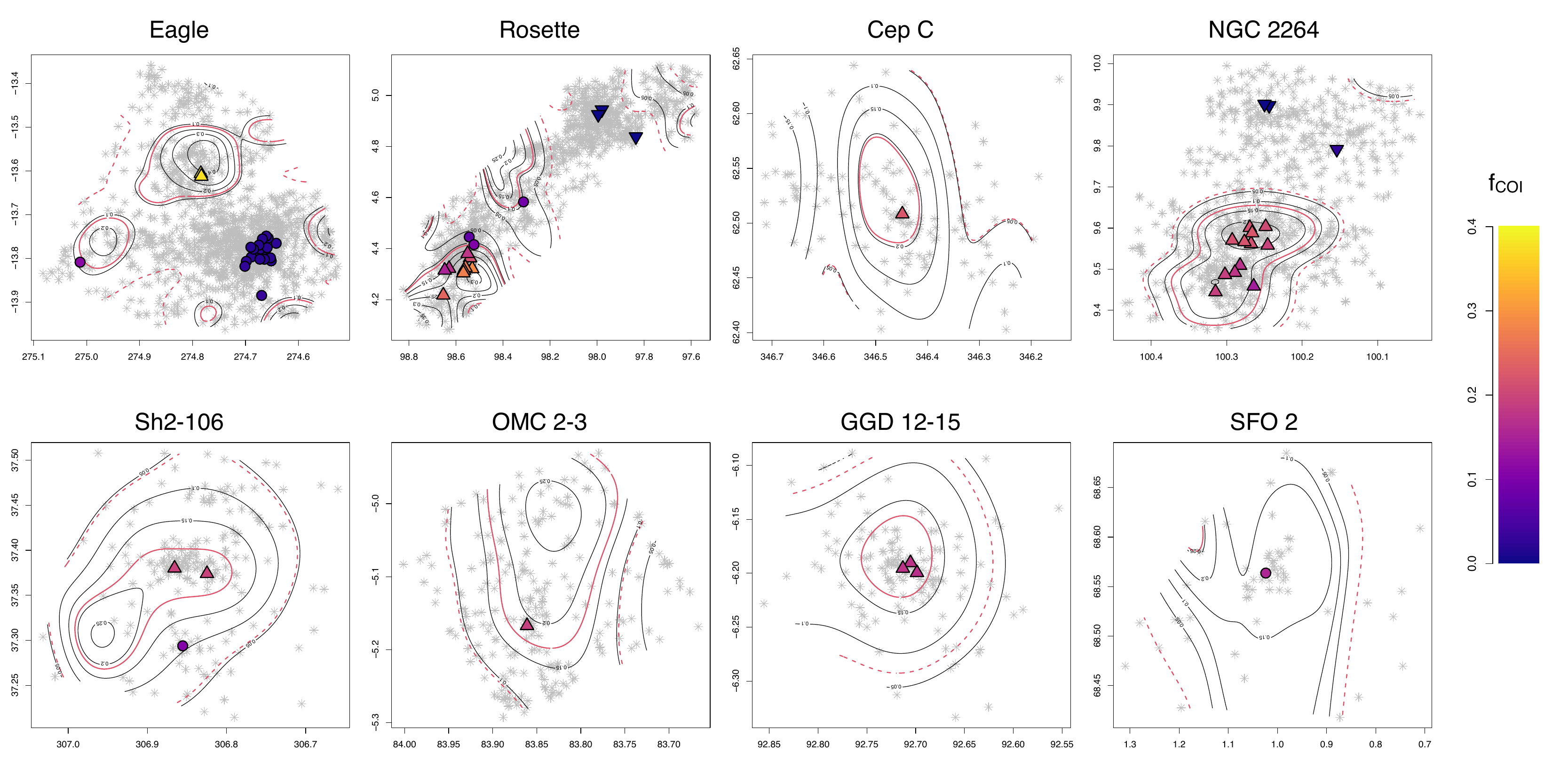}
     \includegraphics[width=0.9\textwidth]{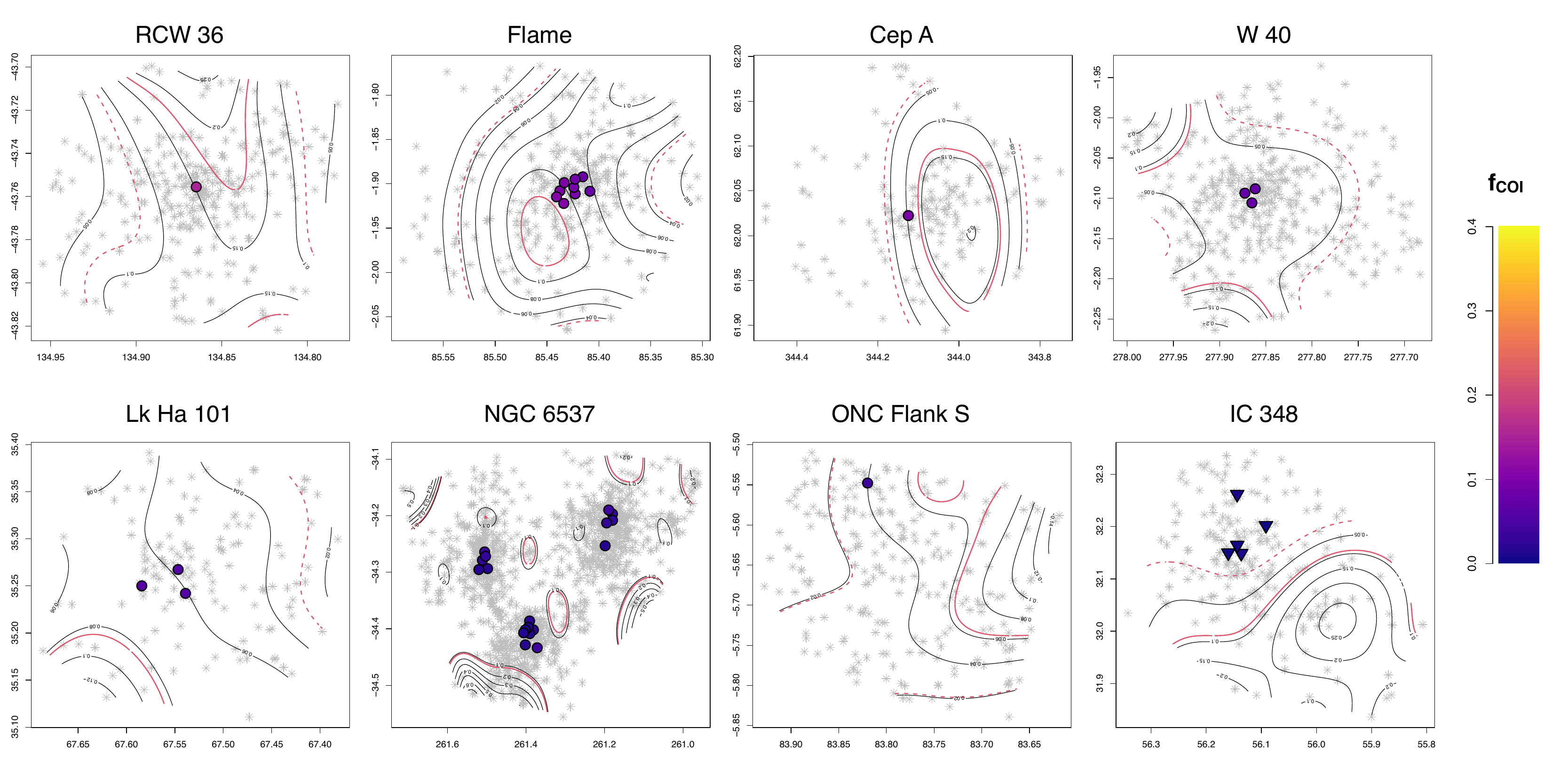}
   
\end{center}
\caption{Maps of IRA regions, with the same symbols as in Figure \ref{mapsHRA}} 
\label{mapsIRA}
\end{figure*}

\begin{enumerate}
    \item {\textbf{Eagle} This HII region in the Serpens constellation, also known as M16, hosts the young cluster NGC 6611 \citet{Stoopetal23}. We find 23 NESTs that contain 14\% of the YSOs in the region. The NEST distribution is dominated by a large group of 19 with average activity, spatially corresponding to NGC 6111 and including a well-populated NEST with $N_{MX}=4.2$. 
    Towards the North of the region, we find two active NESTs, with very high values of $f_{C0/I}\sim 0.38$. West of NGC6111, we find a single, small NEST of average activity but close to the relative risk threshold.  There is also a single small, compact NEST of average activity south of the large group, with similar activity to it. This structure is not coincident with those obtained by \citepalias{Kuhnetal14} or the MST technique (see complementary material).  The relative risk pattern is not correlated with the YSO density, as the largest spatial concentration, NGC6111, is of average activity. Areas of large $f_{C0/I}$ ratios are peripheral, notably overlapping with the eastern and Northern NESTs, but displaced from the southern one. The value of Q=0.8 would point to a lack of substructure, but our results indicate that it is a mixed situation, comprising various small-scale significant substructures and a dominant, large-scale concentration.}
    \item {\textbf{Rosette} The Rosette nebula hosts several young clusters and associations, notably NGC2244. Our analysis finds a rich population of 15 NESTS with significantly different activities and forming various groupings. This is consistent with the low value of Q=0.45, characteristic of substructured distributions. 
Towards the northwest, we find 3 NESTs less active than average, two of them together and corresponding to the known cluster NGC2244. The rest of the NESTs correlate spatially with groupings reported by \citet{Muziketal22}, and also with substructures found by \citepalias{Kuhnetal14} and the MST. In the centre of the image we find a single NEST of average activity, and all the other NESTs are located in the southeast, forming an elongated arrangement. This large grouping shows a significant activity gradient with NESTs of both high and average activity and broadly overlaps high-density gas structures \citep{Hennemannetal10,Motteetal10Hobys}.
The relative risk spatial pattern is complex: in the south, it is mostly consistent with the YSO density, and as we move north-east, the relationship degrades and inverts. In the central area, the higher $f_{C0/I}$ values are displaced from the dense NEST and in the north, the high YSO density region is actually of low relative risk. } 
\item {\textbf{CepC} This region is part of the Cepheus complex, and displays a simple structure consistent with low concentration values ($Q=0.83$). We find a single NEST small, compact, and active, with a high value of $f_{C0/I}=22.5\%$. The relative risk shows a central high $f_{C0/I}$ area and a low-risk NW region, the latter supported by low densities. Our results globally agree with those by \citet{Gutermuthetal11}, who studied the YSO and dust distribution on the whole complex. } 

\item {\textbf{NGC2264} This region in Monoceros hosts a rich, well-characterised, YSO population of different evolutionary stages. As one of the regions with the best studied SF history, we have used it as a benchmark to evaluate the methodology in this work, and more detailed results are avaliable as online complementary material.

The complex has two main large substructures: a more evolved northern one around the massive star S-Mon, and a bilobed active cluster Spokes-Cone \citep[see e.g.][and references therein]{Nonyetal21}. We find a population of 15 NESTs hosting 15\% of the YSOs, consistent with the high level of substructure (Q=0.65) and the general knowledge of the region.  NESTs are organised into three groups: a Northern group corresponding to S-Mon with two evolved NESTs, a single evolved NEST towards the west, and a larger  group of 12 highly active NESTs in Spokes-Cone.  In addition to the general North-South SF gradient of the complex, we detect differences in activity in the Spokes-Cone grouping, with a significant span $\Delta f_{C0/I}$. While all 12 NESTs are active, those in spokes substructure have higher SF than those of the Cone, as reported in \citealt{Venuttietal18}.}%
  
 \item{\textbf{Sh2-106} This is an expanding bipolar HII region in the Cygnus X area \citep{Ballyetal22}. We find three NESTs that account for 19\% of all the objects in the region. Two of them are located towards the center of the region, highly active, and seem to trace the same large-scale structure.  We detect a separate small NEST in the south with average activity, significantly lower than the central ones. The high value of $Q=0.91$ points to the dominance of the central large-scale structure, which also contains a well-populated NEST ($N_{MX}=7.5$). The largest values in $f_{C0/I}$ are shifted east from the density peak traced by the largest NEST, but we are cautious due to the very low YSO density values in the peak. }   
     \item {\textbf{OMC 2-3} This region is part of the Orion Nebula, located between the Orion Nebula Cluster and ONC Flank N, both also part of this work \citep{PetersonMegeath08}. The value of $Q=0.7$ suggests some degree of substructure, but we only find one very small and compact NEST,  active and centered in the field.  The global ratio is $\bar{f}_{C0/I}=13.4\%$ close to the HRA boundary. We note that the Orion fields in these samples divide a known large gas filament.  The YSO density shows larger values towards the southern boundary, where the very dense main Orion Nebula Cluster is located. The relative risk map has a different distribution from the YSOs, with an elongated high $f_{C0/I}$ area vertically aligned with the filament segment in the field. In addition to density gradients, the division of a region, particularly if it cuts a dense structure, can impact the structural properties of the sample, including key parameters for S2D2 such as the reference comparison density. We performed S2D2 in a sample composed of all 4 regions (here referred to as Orion, OMC 2-3, ONC Flank N, and ONC Flank S), to evaluate these effects. We only retrieved a group of NESTs corresponding to the dominant structure, the Orion Nebula Cluster.}
     \item {\textbf{GGD12-15} This is a high mass region, sometimes also identified as IRAS 06084-0611, on the Monoceros R2 complex.  We retrieve a group of three small NESTs with 15\% of the YSO population and high. The group corresponds to the large central structure in the YSO density, which has $Q=1.0$, indicating high concentration levels.  The relative risk map is correlated with the YSO density and has a decreasing outward gradient. Our results are consistent with other work. \citet{Gutermuthetal11} studied YSOs in the complete Monoceros R2 complex field and in connection with the dust estimates, and our NESTs coincide with their large structure, which overlaps with a peak in the dust. Our results are also compatible with \citet{Maaskantetal11}, who report a larger concentration of less evolved objects attributed to sequential SF along the line of sight. }
     \item {\textbf{SFO 2} This region \citep[also known as IRAS 00013+681,7 BRC2, and S171,][]{Getmanetal18} is part of the Cepheus Loop bubble and close to Be 59, also studied in this work. We find a single NEST, small and compact, tracing the central global concentration ($Q=0.88$). The global number of members of this region  { $N=63$ }is low, which can produce high relative density $\rho_{rel}>30$ and large dispersion values. Most of the relative risk map is consistent with the average, with high and low $f_{C0/I}$ values located towards the boundaries, which we note have low densities of YSOs. This young star cluster is located on a bright rimmed cloud, where sequential SF has long been proposed \citep[][]{Sugitanietal95}.}
     \item {\textbf{RCW36} We analyse the young central cluster in this HII region from the Vela C molecular cloud, characterised by gas expansion patterns \citep{Bonneetal22}. We find a single NEST with average activity. The NEST is well populated, with 36 objects corresponding to 12\% of the total YSO sample and a relative population value $N_{MX}=7.2$. Our results are consistent with a global large-scale concentration, also indicated by $Q=0.9$. The relative risk map shows a different pattern from the YSO density: the central concentration is consistent with average,  with high (resp low) $f_{C0/I}$ values towards the north (resp south).}
      \item {\textbf{Flame Nebula} We obtain a group of 9 average NESTs comprising 15\% of YSOs in a concentrated distribution (Q=0.96) consistent with he cluster NGC2024. The group is displaced west of the peak in relative risk, and one of the NESTs is close to its significance threshold. The NESTs are roughly coincident with the main filament \citep{Beslicetal24}, but specific comparisons with the gas density need to be performed. }     
    \item  {\textbf{Cep A} In this portion of the Cepheus complex, we find a single small and compact NEST, with average activity, and consistent with low levels of spatial concentration ($Q=0.84$), similar to Cep C. The relative risk map has a single elongated high-risk peak displaced towards the west from the YSO density peak traced by the NEST, with the eastern part of the field characterised by low $f_{C0/I}$ values. }
    \item {\textbf{W40} This region belongs to the Aquila Rift and is close to Serpens South. We find a group of 3 NESTs containing 7\% of all YSOs compatible with moderate levels of spatial concentration ($Q=0.85$). The large-scale structure traced by NEST is spatially consistent with the highest ranking structure in the hierarchical, large-scale analysis by \citet{Sunetal22}. The relative risk map shows a very different pattern from the YSO density, with average values in the central area and low values on the West. We only find $f_{C0/I}$ values concentrated on the North and South Eastern corners of the field.}

    \item {\textbf{LkHa101} This is an embedded region in the California Molecular Cloud, named by a central Herbig Be star  \citep{Wolketal10, Getmanetal18}. We find three NESTs of average activity, towards the center of the field, consistent with a moderate global large-scale concentration of the YSO sample ($Q=0.84$). The central YSO concentration has an average relative risk, with high $f_{C0/I}$ values towards the Southern area where densities are low.  \citet{Wolketal10} obtained a catalogue for the region and derived a ratio of ClassIII:ClassII: ClassI of 7:5:1, corresponding to a $f_{C0/I}$ ratio of 8\%, larger than but consistent with the values $\bar{f}_{C0/I}=6$\% derived from this sample. }

         \item {\textbf{NGC6357} This SF complex in the Sagittarius arm contains 3 known open clusters of similar sizes and estimated ages of $\sim 1 $ Myr \citep[see e.g.][and references therein]{OrdenesHuanca24}. We obtained  18 NESTs of average activity arranged in three large-scale groupings, compatible with the known substructure of the region and the low values of $Q=0.65$. This is one of two IRA regions where all NESTs have lower $f_{C0/I}$ than average, although the difference is not significant. The values of $\bar{f}_{C0/I}=0.06$ are on the low side of IRA regions, and the uncertainty $\sigma_{f_{C0/I}}$ is slightly larger, so all possible low values of $f_{C0/I}$ are consistent with the average. The relative risk map shows several high-risk peaks, small and decoupled from the YSO density. Our results are globally consistent with the general knowledge of the complex, host of several bubbles and massive stars. }
    \item {\textbf{ONC Flank S}: This is another part of the Orion Nebula, with a situation similar to the previously described in OMC2-3. We find a single, small NEST towards the ONC, located North of the observed field. We note that while the number of stars and structural characteristics of this sample are similar to the previous one, the global ratio is very different $\bar{f}_{C0/I}=4\%$, close to the lower boundary of the IRA regions. The relative risk map is different from the complete YSO distribution, with an East-to-West increasing gradient.}
    \item      {\textbf{IC348} This region, part of the Perseus complex and close to NGC1333, has a group of 5 NESTs towards the North. IC348 was also analysed in \citetalias{Gonzalezetal21} but using the sample provided by \citet{Luhmanetal16}. The two samples have different spatial coverage and number of stars: the one used in this work has fewer stars and a smaller field. This sample has a northern NEST that was not retrieved by our previous analysis, and, conversely, in \citetalias{Gonzalezetal21} we found a separate NEST towards the South, but the southernmost group of three NESTs and the western one coincide. This is the only IRA region where all retrieved NESTs  {have significantly low activity}. The ratio of Class 0/I objects we obtain, $f_{C0/I}=0.06$, is compatible with that of \citet{Young2015}, 0.08, but their ratio of Class II objects, 0.73, is much larger than the one in this work $f_{CII}=0.31$ . Our estimate of class II members is closer to that of \citet{Luhmanetal16} $\sim 0.4$. The relative risk map shows an increasing N-S gradient. Our global results are consistent with the Perseus analysis in \citet{Gutermuthetal11}, which shows a Northern concentration of YSOs displaced from the southern peak in dust density. }

  \end{enumerate}

\subsection{LRA regions}

\begin{figure*}[ht!]
\begin{center}
\includegraphics[width=0.9\textwidth]{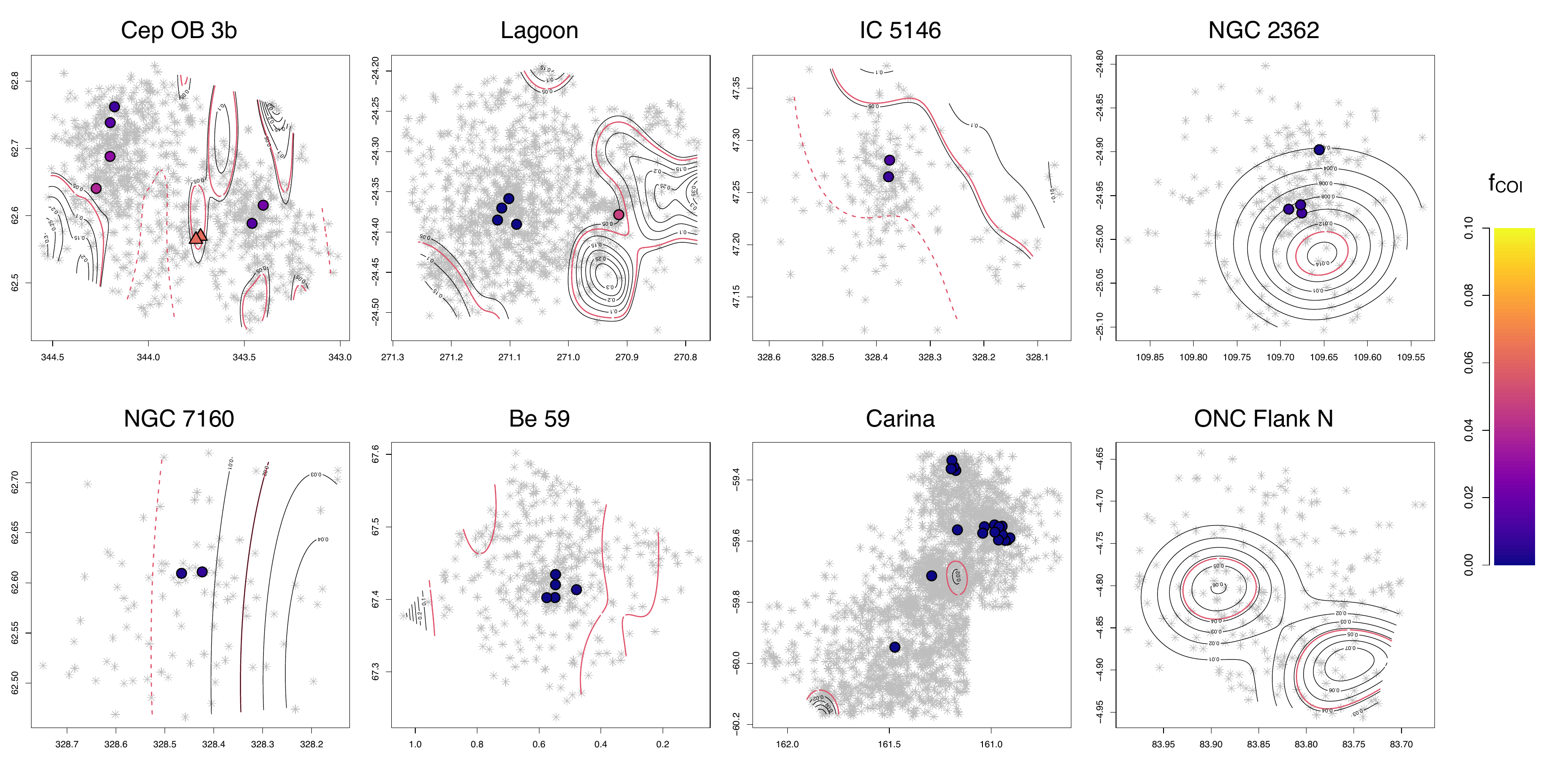}
			
\end{center}
\caption{Maps of LRA regions with NESTs, with the same symbols as in Figure \ref{mapsHRA}}
\label{mapsLRA}
\end{figure*}

\begin{enumerate}
    \item {\textbf{Cep OB 3b} This is a well-populated Cepheus region in the late stages of gas dispersal \citep{Karnathetal19}, consistently with a low global ratio of Class 0/I objects. We find 8 small NESTs that only account for 6\% of the YSOs in the region. NESTs are grouped in three distinct areas, compatible with previously retrieved substructures, and also with the low $Q=0.65$ values.  Relative risk has a globally complex pattern, with peaks decoupled from the global YSO density.\citet{Gutermuthetal11} and \citet{Allenetal12} also show a YSO distribution different from that of the gas, with the large eastern and western subclusters only partially overlapping high-density gas. Our results are consistent with residual recent SF associated with gas remnants, and the general knowledge of the region. Additionally, if we use the ratio of objects of Class II $f_{CII}$ as a youth indicator, we find significantly larger values in the western subcluster, in agreement with the fractions of objects with disks found by \citet{Allenetal12}.}
    \item {\textbf{Lagoon} Lagoon Nebula, also known as M8, hosts an HII region and open cluster NGC6530.  We find 5 average NESTs in the region, organised into two separate groups. The eastern group has 4 NESTs, and the western one is a well-populated NEST with $N_{MX}=6$. The different relative weights of both substructures can explain the value of $Q=0.8$, indicative of CSR in the box-fractal/radial paradigm. Despite all NESTs being compatible with average, we find significant differences in the activity of both subclusters, $\Delta f_{NEST}>\sigma_{{f}_{C0/I}}$, with the western subcluster significantly more active than the eastern. Our results are compatible with recent spatiokinematical structures retrieved by \citet{Jiaetal24}. In particular, the western structure would correspond to sub-3, recently formed according to their kinematic tracing analysis, and the eastern clump spatially overlaps with their main substructure, sub-1. \citet{Jiaetal24} predict that substructure 3 will merge with 1, but the fate of other substructures is unclear, as the picture is very complex.  Such a complex scenario agrees with the previous spatiokinematical analysis by \citet{WrightParker19}. Their results indicate a strong dynamical evolution of the region, which would have undergone cold collapse from a more extended and substructured distribution.}
    \item {\textbf{IC5146} This is a young open cluster in the Cocoon nebula, characterised by an HII region surrounded by filamentary gas structures that are currently forming dense cores \citep[][]{Chungetal24}. While this hints to future SF in the region, the current YSO sample we analyse has very low values of $f_{C0/I}=0.03$.
    We find 2 NESTs, tracing the large-scale central concentration ($Q=0.86$). Both NESTs are of average activity, and the Southern one is well populated with $N_{MX}=6.4$. The relative risk shows a global decreasing trend from North to South, but we need to remember that $f_{C0/I}$ values are low and the periphery has relatively low YSO density. The Class II relative risk map shows two large-scale structures with significantly high risk. One of them is towards the lower density western structure, and the main one coincides with the YSO concentration, with one NEST showing a significant Class II ratio and the other consistent with average but close to the boundary.} 
    \item {\textbf{NGC2362} This cluster is characterised by its low extinction and a rich population with several massive stars \citep{Damianetal21}. We find 4 NESTs with average recent activity on a sample that shows signs of a strong global concentration ($Q=0.93$). Three of the NESTs are grouped towards the centre, while a single one appears on the outskirts towards the North. Regarding the evolutionary stage of the cluster, the classified sample is globally very evolved, with an extremely low ratio of Class 0/I objects (there is only one) and less than 10\% of Class II. The relative risk of Class II is mostly consistent with the average, as high/low $f_{CII}$ values are on peripheral, low-density areas. }
    \item {\textbf{NGC7160} This open cluster in the Cepheus OB2 region is relatively old  \citep[estimates compiled by][indicate an age between 8 and 15 Myr]{Mendigutiaetal22}, and we accordingly find a very low $f_{C0/I}=0.01$ (with only one source), and also a very low ratio of Class II objects $f_{CII}=0.05$. We note that while the region has a low global number of members $N=96$, almost all of them are classified as stellar. We retrieve two small NESTs of average activity located towards the center of the region tracing the large-scale overdensity ($Q=0.9$). The Class II relative risk map shows a descending patterns towards the East, with larger values on the south. } 
    \item {\textbf{Be 59} Berkeley 59 is the main cluster within an HII region of Cepheus OB4, close to SFO2. We find 5 NESTs of average activity in a large group consistent with a large-scale concentration, also indicated by $Q=0.86$. The central part of the relative risk map is consistent with average with high areas towards the west and on the easter boundaries. The distribution of Class II sources is more complex, with several areas of both high and low activity surrounding the central area which is consistent with average.  These results are broadly consistent with the triggered SF scenario explored by \citet{Mintzetal21}, with the whole Cep OB4 complex having a low ratio of Class I/II objects in the cluster and increasing towards the outskirts of the extended structure.}
    \item {\textbf{Carina} As previously mentioned, this region was already analysed in \citetalias{Gonzalezetal21} where we used a physical 1000AU merging limit, while here the physical distance associated with the constant $2 ^{"}$ resolution limit corresponds to 4.6 kAU. The results are very similar, with 10\% of the members grouped in 19 NESTs that trace most of the known open clusters in the region. Trumpler 14 is still the dominant structure in the sample, traced by a group of NESTs that includes the most populated one ($N_{MX}=25.60$). 
    There were only 2 members of the region classified as Class 0/I, meaning extremely low values of $f_{C0/I}=0.001$, so all  NESTs (and actually most of the region) have $f_{C0/I}$ values consistent with average. The value $f_{CII}=0.1$ is also relatively low, and the Class II relative risk map has peaks all over the region, although none coincide with the position of NESTs.
    While the general picture of the region from this sample points to very low recent SF activity, other studies show relevant age spans in Carina and very young age estimates for Trumpler 14 \citep[see e.g.][ and references therein]{Itrichetal24}, which could point to some recent SF in the region undetected in this study. }
    \item {\textbf{ONC Flank N}: This field is a peripheral part of the Orion Molecular Cloud where we did not retrieve any significant small-scale substructure. The distribution generally agrees with CSR, as pointed by $Q=0.78$ and the high values of $\rho_{rel}$. This region corresponds to the peripheral part of the Orion molecular cloud filament, so the previously mentioned caveats regarding the window selection hold. The $f_{C0/I}$ map shows two high-risk areas, but we must note that they are only traced by the few Class 0/I members of the region. The class II relative risk map, with a large support,  shows significantly high values on the southern boundary, towards the large gas filament.  }

\end{enumerate}

\subsection{Flagged regions}

\begin{figure}[ht!]
\begin{center}
\includegraphics[width=\columnwidth]{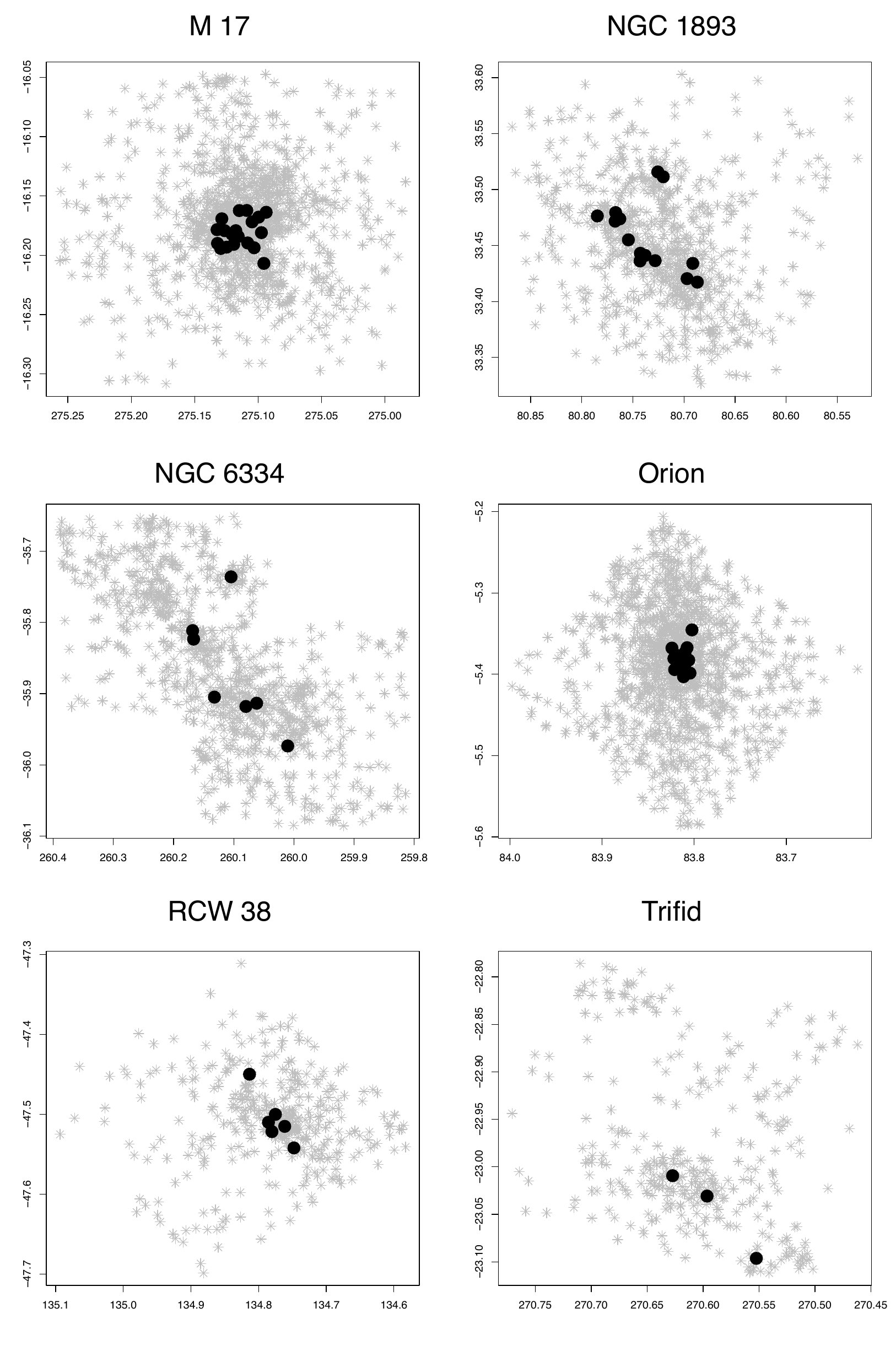}
			
\end{center}
\caption{Maps of regions with a classified subsample flagged as non representative. Grey dots represent the YSOs in the area, and black markers show the position of the centroids of NESTs.}
\label{mapsFlag}
\end{figure}

Most of these regions are very embedded, high-mass regions, where the presence of gas and massive stars hinders observations. Indeed, a substantial fraction of members in these samples do not have the required IR photometric information at all, or with insufficient quality for classification \citepalias{Gutermuthetal09}.

\begin{enumerate}
    \item     {\textbf{Orion} Main cluster in the Orion cloud, very well studied as one of the closest examples of active high mass SF. Peripheral sections of this same cloud have been previously analysed as part of this work (namely ONC Flank N, ONC Flank S, and OMC 2-3). The cluster is highly embedded, and 55$\%$ of the objects in our sample remained unclassified due to poor photometry. We retrieve a group of 13 NESTs on the centre, tracing a single large-scale concentration. The structural simplicity of this region is supported by the value $Q=0.87$ and the presence of well-populated NESTs ($N_{MX}=5.80$). Some authors have long observed age / evolutionary outward gradient from the filament \citep[e.g.][]{Getmanetal18AgeGradients, StutzGould16}. 
    Despite the attention received by this region, there are still open questions, including debate on global formation scenarios and whether SF was continuous or episodic \citep[e.g.][]{Alzateetal23, Kounkeletal22,Beccarietal17}.}
    \item {\textbf{M17} This is a very dynamic and active region, host of a large population of massive stars and the very young, well-populated, open cluster NGC 6618.  Due to the associated observational difficulties, $86\%$ of the members in this sample could not be classified due to the lack of photometry of enough quality. We find a group of 19 NESTS containing 18\% of the YSO sample that traces a high-density large-scale YSO concentration ($Q=1.01$). The history of the cluster is still under discussion, particularly the role of a nearby OB group as the initial trigger of SF in the area \citep[see][and references therein]{Stoopetal24}.}
    \item {\textbf{RCW38} RCW 38 is known as one of the closest high-mass star-forming regions. In this YSO sample, we could not classify $59\%$ of the members due to photometric quality issues. We retrieve a total of 6 NESTs, 5 of which trace the single, large-scale central YSO overdensity, supported by $Q=0.91$ and $N_{MX}=26.75$. We also find a separate single, small and compact NEST towards the North. \citet{Fukuietal16} proposed cloud-cloud collision as the trigger of the formation of massive stars in the region.}
    \item {\textbf{NGC1893} NGC 1893 is a young open cluster embedded in an HII region \citep{Damianetal21}. Despite its moderate extinction, it is the most distant, $d=3.6$ Kpc and $44\%$ of its members did not have sufficient photometric information for classification. It is the only region with a relevant fraction of objects classified as non-stellar, 11\%. There is clear large-scale overdensity in the region, also supported by $Q=0.92$, and our results suggest some level of substructure within it. We retrieved 14 NESTs that contain 12\% of the members of the region. 12 NESTs are visually aligned with a large-scale elongated density structure, while the other two form a distinct group towards the north. }
    \item {\textbf{Trifid:} Trifid is a bright HII nebula in Sagittarius, also known as M20, host of open cluster NGC6514. $53\%$ of the sample did not meet quality requirements.  We find three small NESTs containing 9\% of the YSOs, consistent with a substructured distribution (also suggested by $Q=0.69$). The two central NESTs correspond to the main cluster, and the southern NEST is located on a filament cut by the field boundary \citepalias[as indicated in][]{Kuhnetal14}. In this region, cloud-cloud collision has also been proposed as star formation mechanism \citep{Toriietal17}, although recent kinematical studies of YSOs propose that expansion of an HII region \citep{Kuhnetal22} could have triggered star formation. }
    \item {\textbf{NGC6334} This is a high-mass complex that contains HII regions and high-density ridge, where $60\%$ of the sources lacked the required photometric information. The region is highly active and contains a considerable population of dense cores with complex spatial distribution and various evolutionary stages which was attributed to hierarchical fragmentation by \citet{Sadaghianietal20}. We find 7 small, compact NESTS that only contain 6\% of all the YSOs dispersed over a large portion the region, which has a value of $Q=0.71$, pointing to a highly substructured distribution. The positions of NESTs seems close to the hubs and ridges identified in the molecular cloud \citep[studied in][]{Russeiletal13,Tigeetal17}. 
    At this stage, the evidence is consistent with most hierarchical SF scenarios, and further analyses considering the molecular cloud and overlaps between NESTs and the cloud structures   {would be interesting. }}
\end{enumerate}

\end{appendix}
\end{document}